# Optimized Energy Aware 5G Network Function Virtualization

Ahmed N. Al-Quzweeni, Ahmed Q. Lawey, Taisir E.H. Elgorashi, and Jaafar M.H. Elmirghani

*Abstract*—In this paper, network function virtualization (NVF) is identified as a promising key technology that can contribute to energy-efficiency improvement in 5G networks. An optical network supported architecture is proposed and investigated in this work to provide the wired infrastructure needed in 5G networks and to support NFV towards an energy efficient 5G network. In this architecture the mobile core network functions as well as baseband function are virtualized and provided as VMs. The impact of the total number of active users in the network, backhaul/fronthaul configurations and VM inter-traffic are investigated. A mixed integer linear programming (MILP) optimization model is developed with the objective of minimizing the total power consumption by optimizing the VMs location and VMs servers' utilization. The MILP model results show that virtualization can result in up to 38% (average 34%) energy saving. The results also reveal how the total number of active users affects the baseband VMs optimal distribution whilst the core network VMs distribution is affected mainly by the inter-traffic between the VMs. For real-time implementation, two heuristics are developed, an Energy Efficient NFV without CNVMs inter-traffic (EENFVnoITr) heuristic and an Energy Efficient NFV with CNVMs inter-traffic (EENFVwithITr) heuristic, both produce comparable results to the optimal MILP results.

*Index Terms*—. 5G networks, Backhaul, BBU, Energy Efficiency, Fronthaul, IP over WDM, MILP, Network Function Virtualization, NFV.

## I. INTRODUCTION

According to Cisco Visual Networking Index, mobile data traffic will witness seven pleats between 2016 and 2021 and will grow at a Compound Annual Growth Rate (CAGR) of 46% reaching 48.3 exabytes per month by 2021 [2]. This growth is driven by a number of factors such as the enormous amount of connected devices and the development of data-greedy applications [3]. With such a tremendous amount of data traffic, a revolutionary mobile network architecture is needed. Such a network (5G) will contain a mix of a multiple access technologies supported by a significant amount of new spectrum to provide different services to a massive number of different types of users (eg. IoT, personal, industrial) at high data rate, any time with potentially less than 1 ms latency [4]. 5G networks are expected to be operational by 2020 where a huge number of devices and application will use it [5].

Users, applications, and devices of different kinds and purposes need to send and access data from distributed and centralized servers and databases using public and/or private networks and clouds. To support these requirements, 5G mobile networks have to possess intelligence, flexible traffic management, adaptive bandwidth assignment, and at the forefront of these traits is energy efficiency. Information and Communication Technology (ICT) including services and devices are responsible for about 8% of the total world energy consumption [6] and contributed about 2% of the global carbon emissions [7]. It is estimated that, if the current trends continue, the ICT energy consumption will reach about 14% of the total worldwide consumption by 2020 [6]

There have also been various efforts from researchers on reducing the power consumption in 5G networks. For instance, the authors in [8] focused in their work on the power consumption of base stations. They proposed a time-triggered sleep mode for future base stations in order to reduce the power consumption. The authors in [9] investigated the base stations computation power and compared it to the transmission power. They concluded that the base station computation power will play an important role in 5G energy-efficiency. The authors of [10] developed an analytical model to address the planning and the dimensioning of 5G Cloud RAN (C-RAN) and compared it to the traditional RAN. They showed that C-RAN can improve the 5G energy-efficiency. The research carried out in [11] focused on offloading the network traffic to the mobile edge to improve the energy-efficiency of 5G mobile networks. The authors developed an offloading mechanism for mobile edge computing in 5G where both file transmission and task computation were considered.

Virtualization has been proposed as an enabler for the optimum use of network resources, scalability, and agility. In [12] the authors stated that NFV is the most important recent advance in mobile networks where among its key benefits is the agile provisioning of mobile functions on demand. The fact that it is now possible to separate the functions form their underlying hardware and transfer them into software-based mobile functions as well as provide them on demand, presents opportunities for optimizing the physical resources and improving the network energy efficiency.

In this paper, network function virtualization is identified as a

Manuscript received MONTH DD, YYYY; revised MONTH DD, YYYY and MONTH DD, YYYY; accepted MONTH DD, YYYY. Date of publication MONTH DD, YYYY; date of current version MONTH DD, YYYY. This work was supported by the Engineering and Physical Sciences Research Council (ESPRC), INTERNET (EP/H040536/1), and STAR (EP/K016873/1).

The authors are with the School of Electronic and Electrical Engineering, University of Leeds, Leeds LS2 9JT, U.K

promising key technology that can contribute to the energy-efficiency improvement in 5G networks. In addition, an optical network architecture is proposed and investigated in this paper to provide the wired infrastructural needed in 5G networks, and to support NFV and content caching. In the literature, NFV was investigated either in mobile core networks [13-15] or in the radio access network [16-18] of the mobile network and mostly using pooling of resources such as the work in [19, 20]. In contrast, virtualization in this paper is not limited to a certain part in the mobile network, but is applied in both the mobile core network and the radio access network. Moreover, it is not confined to pooling the network resources, but is concerned with mobile functions-hardware decoupling and considers converting these functions into software-based functions that can be placed optimally. A Mixed Integer Linear Programming model and real-time heuristics are developed in this paper with the goal of improving the energy-efficiency in 5G mobile networks.

## II. NFV IN 5G NETWORKS

According to the third generation partnership project (3GPP) the evolved packed core (EPC) is an important step change [21]. There are four main functions in the EPC [22, 23] illustrated in Fig. 1: the packet data network gateway (PGW), the serving gateway (SGW), the mobility and management entity (MME), and the policy control and charging role function (PCRF).

The work in this paper extends our work in [24, 25] to include a number of factors such as the total number of active users in the network during the day, the backhaul and fronthaul configuration and the required workload for baseband processing. It introduces an optical-based framework for energy efficient NFV deployment in 5G networks and provides full MILP details and associated heuristics. In this framework, the functions of the four entities of mobile core network are virtualized and provided as one virtual machine, which is dubbed "core network virtual machine" (CNVM). For the radio access side, the BBU and RRU are split and the function of the BBU is virtualized and provisioned as a "BBU virtual machine" (BBUVM). Consequently, the wireless access network of the mobile system will encompass only the RRU that remain after the RRU-BBU decoupling. RRU is referred to here as "RRH" (as in a number of studies [26-28]) after it is separated from BBU. The traffic from CNVM to RRH is compelled to pass through BBUVMs for baseband processing, as in Fig. 2. Moreover, the capabilities of Passive Optical Networks (PON) are leveraged as an energy-efficient broadband access network to connect the IP over WDM core network to RRH nodes, and to represent the wired access network of our proposed system. Fig. 3 shows three locations that can accommodate virtual machines (VMs) of any type (BBUVMs or CNVMs), which are the optical network unit (ONU), optical line terminator (OLT), and the IP over WDM nodes. For simplicity, the nodes where the hosted servers are accommodated are referred to as "Hosting Nodes".

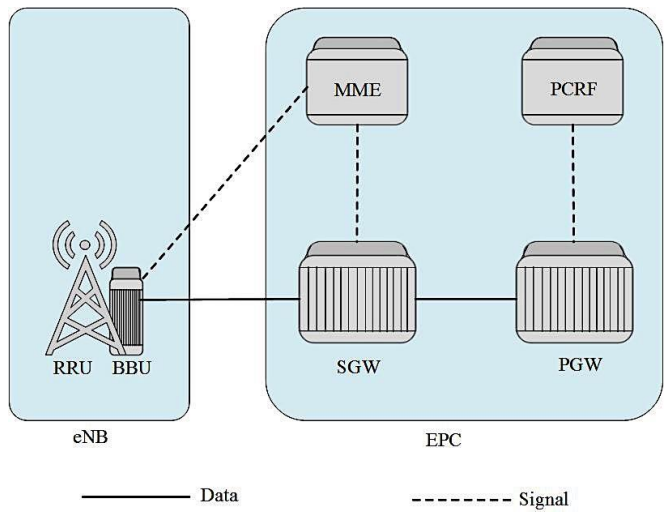

Fig. 1. Evolved Packet System Architecture

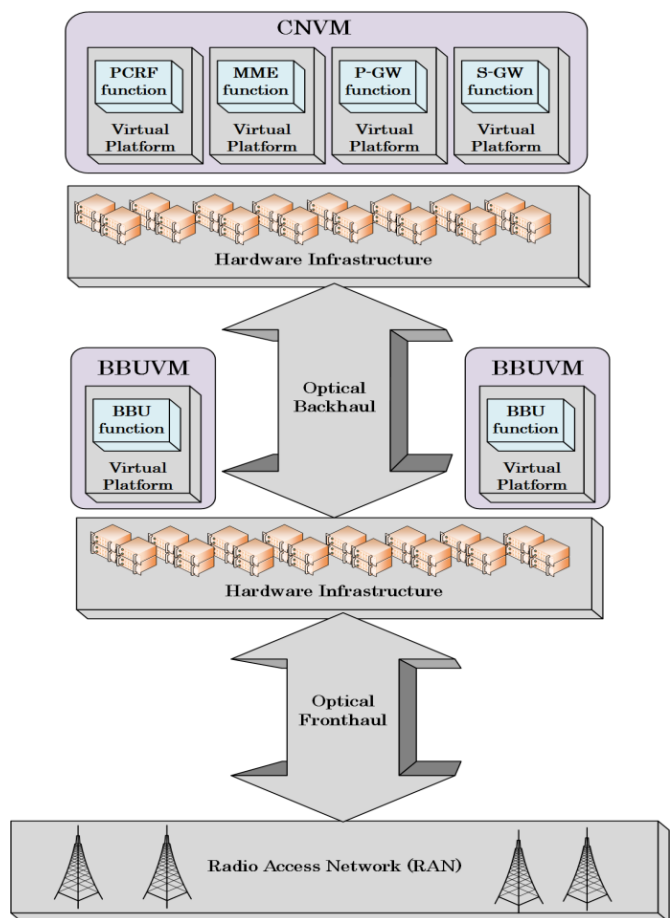

Fig. 2. The proposed architecture for Energy Efficient NFV in 5G

The hosting nodes (ONU, OLT and IP over WDM nodes) might host one VM or more than one VM of the same or different types, bringing forth the creation of small clouds, or "Cloudlets". Therefore, the proposed architecture will provide an agile allotment of services and processes through flexible distribution of VMs over the optical network (PON and IP over WDM network), which is one of the main concerns of this work in minimizing the total power consumption. Based on this architecture, a MILP formulation has been developed with the overall aim of minimizing power consumption.

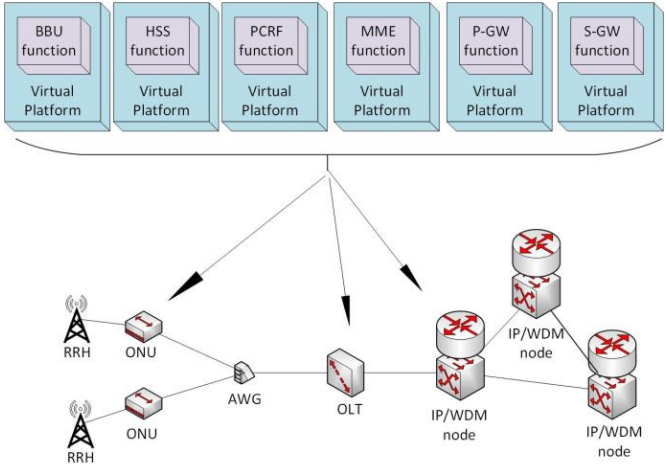

Fig. 3. Evolved Packet System Architecture

## III. FRONTHAUL AND BACKHAUL CONFIGURATION AND THE AMOUNT OF BASEBAND PROCESSING WORKLOAD

This section illustrates the configuration of the fronthaul and backhaul used in the proposed network; so that the ratio of the backhaul to the fronthaul data rate could be calculated. Fronthaul is the network segment that connects the remote radio head (RRH) to the baseband unit (BBU) [29], whilst the network segment that connects the BBU to the mobile core network (CN) is called "backhaul" [30]. The internal interface of the fronthaul is defined as a result of the digitization of the radio signal according to a number of specifications. The well-known and most used specification among radio access network (RAN) vendors is the Common Public Radio Interface (CPRI) specification [31] which is implemented using digital radio over fiber (D-RoF) techniques. On the other hand, the backhaul interface leverages Ethernet networks as they are the most cost effective network for transporting the backhaul IP packets [32, 33].

In order to adequately determine the data rate in each network segment (backhaul and fronthaul), we will start with the physical layer of the current mobile network which is the Long-Term Evolution (LTE) network. The LTE network uses single-carrier frequency-division multiple access (SC-FDMA) uplink (UL), whilst orthogonal frequency-division multiple access (OFDM) is used in the downlink (DL) [34]. In both techniques, the transmitted data are turbo coded and modulated using one of the following modulation formats: QPSK, 16QAM, or 64QAM with 15 kHz subcarriers spacing [35]. A generic frame is defined in LTE which has 10 ms duration and 10 equal-sized subframes. Each subframe is divided into two slot periods of 0.5 ms duration [36]. Depending on the cyclic prefix (CP) used, slots in OFDMA have either 7 symbols for normal CP or 6 symbols for extended CP [37]. Fig. 4 illustrates an LTE downlink frame with normal CP. In the LTE frames, a resource element (RE) is the smallest modulation structure which has one subcarrier of 15 kHz by one symbol [38]. Resource elements are grouped into a physical resource block (PRB) which has dimensions of 12 consecutive subcarriers by one slot (6 or 7 symbols). Therefore, one PRB has a bandwidth of 180 kHz (12 × 15 kHz). Different transmission bandwidths use different number of physical resource blocks (PRBs) per time slot (0.5 ms) which are defined by 3GPP [39]. Fig. 5 illustrates the LTE downlink resource grid. For instance, 10 MHz transmission bandwidth has 50 PRBs whilst 20 MHz has 100 PRBs [40]. If 10 MHz bandwidth is used with 16QAM (6 bits/symbol) and 7 OFDM symbols (short CP), we have

$$\frac{\left(50\ RB \times \frac{12\ subcarriers}{RB} \times \frac{7\ symbols}{subcarrier} \times \frac{6\ bits\ (QAM)}{symbol}\right)}{0.5\ ms\ time\ slot} \quad (1)$$

$$= 50.4\ Mbps$$

$$50.4\ Mbps\ \times 0.874\ system\ efficiency = 44.0496\ Mbps \quad (2)$$

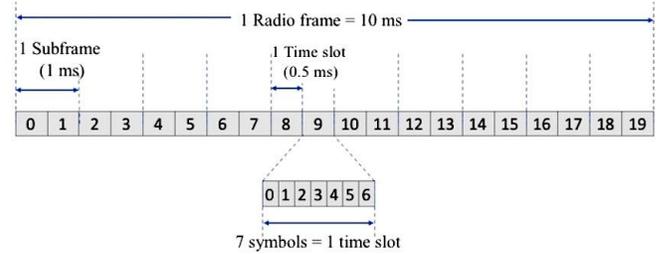

Fig. 4. LTE downlink frame with normal CP

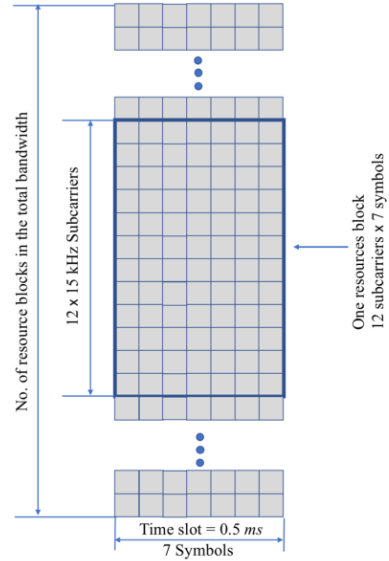

Fig. 5. LTE downlink resources grid

It is worth mentioning that for each transmission antenna, there is one resource grid (50 PRBs for 10 MHz); therefore in $2 \times 2$ MIMO the previous data rate is doubled (100.8 Mbps) [41].

The transmission of user plane data is achieved in the form of In-phase and quadrature (IQ) components that are sent via one CPRI physical link where each IQ data flow represents the data of one carrier for one antenna that is called Antenna-Carrier (AxC) [42]. A number of parameters affect the data carried by AxC, [41]:

Sampling frequency which is calculated as: subcarrier BW (15 kHz) times the FFT window (size). The FFT size is chosen to be the least multiple of 2 that is greater than the ratio of the radio signal bandwidth to the subcarrier BW. For instance, if the radio bandwidth is 10 MHz, the FFT size is the least multiple of 2 that is greater than 666.67 (10 MHz / 15 kHz) which is 1028 ($2^{10}$). In this case the sampling frequency is calculated as $150\ kHz \times 1024 = 15.36\ MHz$. Using the same approach, the sampling frequency at 20 MHz radio bandwidth system is 30.72 MHz.

IQ sample width (*M*-bits per sample): According to the CPRI specification, the *IQ* sample width supported by CPRI is between 4 and 20 bits per sample for *I* and *Q* in the uplink and it is between 8 and 20 in the downlink [42]. For instance, with *M* = 15 bits per sample; one AxC contains 15 bits per sample for *I* and 15 bits per sample for *Q* which are 30 ($2 \times M$) bits per sample *I* and *Q* which are transported in sequence: $I_0Q_0I_1Q_1 \ldots I_{14}Q_{14}$. The *IQ* sample data rate can be calculated by multiplying the number of bits per sample by the sampling frequency. For instance; for a radio bandwidth of 10 MHz ($f_s$ =15.36 MHz) and *IQ* samples 15 (*M* = 15) the *IQ* data rate is:

$$(2 \times M) \times f_s = (30\ bits/sample) \times 15.36\ MHz \quad (3)$$
$$= 0.4608\ Gbps$$

CPRI data rate is designed based on the Universal Mobile Telecommunications System (UMTS) chip rate [42] which is 3.84 Mbps [43, 44]. Therefore, one basic CPRI frame is created every Tc= 260.416 ns (1/3.84 MHz) and this duration should remain constant for all CPRI options and data rates. According to CPRI specification in [42], one basic CPRI frame consists of 16 words indexed (*W*=0…15), where the first word is reserved for control. The length of the frame word (*T*) depends on the CPRI line rate as specified by CPRI specification in [42]. Accordingly, the transmission of *AxC* data will be expanded by a factor of 16/15 (15 bits payload, 1 bit control and management). In addition to the sampling rate $f_s$ that is calculated earlier, *AxC* data needs to be coded using either 8B/10B or 64B/66B.

To put all these calculations together, let's start with the number of bits per word in the CPRI frame. The number of bits per word is equal to the total number of bits per frame divided by the frame payload words (15 words). Recall that the frame duration should be constants (206.416 ns); therefore:

$$\frac{(no\ of\ bits\ per\ word\ \times\ 15\ words)}{samples\ of\ IQ\ f_{IQ}} = 260.416\ ns \quad (4)$$

$$no\ of\ bits\ per\ word\ (Nbpw) = \frac{f_{IQ} \times 260.416\ ns}{15}. \quad (5)$$

one CPRI frame word has *N*bpw bits, since the CPRI frame has 16 words:

$$NbpF = Nbpw \times (15\ payload\ words) \\ + Nbpw \times 1\ control\ word \quad (6)$$

$$NbpF = Nbpw \times (15 + 1) = \frac{f_{IQ} \times 260.416\ ns}{15} \times 16 \quad (7)$$

to calculate the data rate in one CPRI frame

$$\frac{NbpF}{260.416\ ns} = \frac{\frac{f_{IQ} \times 260.416\ ns}{15} \times 16}{260.416\ ns} = f_{IQ} \times \frac{16}{15} \quad (8)$$

by replacing $f_{IQ}$ with $2 \times M \times f_s$
where *M* is defined earlier as the number of *IQ* bits.
In addition, AxC data are coded by either 8B/10B or 64B/66B. By putting these together, the CPRI data rate is calculated as

$$2 \times M \times f_s \times \frac{16}{15} \times L_{coding}. \quad (9)$$

Finally, the ratio of the backhaul to fronthaul data rate is calculated as:

$$\frac{backhaul\ (IP)data\ rate}{fronthaul\ (CPRI)data\ rate} \\ = \frac{44.0496\ Mbps}{327.68\ Mbps} \times 100\ \% = 13.44\ \% \quad (10)$$

Therefore, depending on coding, sampling, quantization, and other parameters; the baseband processing adds overheads to the backhaul traffic as it passes through the BBU. In this work the ratio (13.44%) calculated in (10) is used in our model, whilst the amount of workload in Giga Operation Per Second (GOPS) needed to process one user traffic is used based on the following relation which is explained in [45]:

$$wl = \left(30 \cdot A + 10 \cdot A^2 + 20\frac{M}{6} \cdot C \cdot L\right) \cdot \frac{R}{50} \quad (11)$$

where:
*wl*: is the baseband workload in (GOPS) needed to process one user traffic,
*A*: number of antennas used,
*M*: modulation bits,
*C*: the code rate,
*L*: number of MIMO layers
*R*: number of physical resource blocks allocated for the user.

IV. MILP MODEL

This section introduces the MILP model that has been developed to minimize the power consumption due to both processing by virtual machines (hosting servers) and the traffic flow through the network. As mentioned in the previous section, the MILP model considers an optical-based architecture with two types of VMs (BBUVM and CNVMs) that could be accommodated in ONU, OLT and/or IP over WDM as in Fig. 6. The maximum number of VM-hosting servers considered was 1, 5, and 20 in ONU, OLT, and IP over WDM nodes respectively, which is commensurate with the node size and its potential location and hence space limitations (together with the size of exemplar network considered in the MILP). All VM-hosting servers were considered as sleep-capable servers for the purpose of VM consolidation (bin packing)

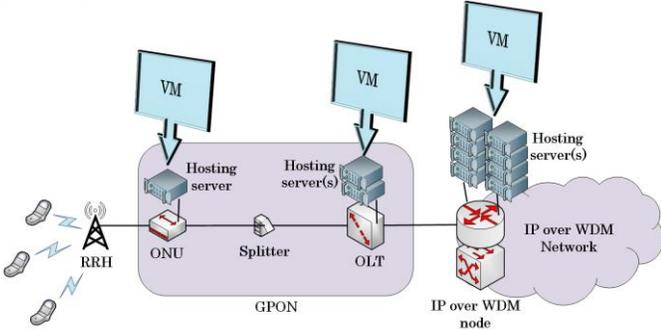

Fig. 6. Candidate locations for hosting VMs in the proposed architecture

For a given request, the MILP model responds by selecting the optimum number of virtual machines and their location so that the total power consumption is minimized.

The following indices, parameters, and variables are defined to represent the developed model:

TABLE I
ENERGY EFFICIENT NFV MILP MODEL INDICES

| Indices | Comments |
|---|---|
| $x, y$ | Indices of any two nodes in the proposed model |
| $m, n$ | Indices of any two nodes in the physical layer of the IP over WDM network |
| $i, j$ | Indices of any two nodes in the IP layer of the IP over WDM network. |
| $r$ | Index of RRH node |
| $h, u, p, q$ | Indices of the nodes where the VM could be hosted |

TABLE II
ENERGY-EFFICIENT NFV MILP MODEL PARAMETERS

| Parameters | Comments |
|---|---|
| $R$ | Set of RRH nodes |
| $U$ | Set of ONU nodes |
| $L$ | Set of OLT nodes |
| $N$ | Set of IP over WDM nodes |
| $T$ | Set of all nodes (RRH, ONU, OLT, and IP over WDM nodes) |
| $NN_m$ | Set of neighbors of node $m$ in the IP over WDM network, $\forall m \in N$ |
| $TN_x$ | Set of neighbors of node $x$, $\forall x \in T$ |
| $H$ | Set of hosting nodes (ONU, OLT, and IP over WDM nodes) |
| $l$ | Line coding rate (bits per sample) |
| $y$ | Number of MIMO layers (ie number of data streams) |
| $q$ | Number of bits used in QAM modulation |
| $a$ | Number of antennas in a cell |
| $cp$ | CPRI link data rate |
| $\Psi X$ | Maximum BBU workload needed for fully loaded RRH (GOPS); calculated as: $30 \cdot a + 10 \cdot a^2 + 20 \cdot q \cdot l \cdot y$ |
| $\Psi S$ | Server CPU maximum workload (GOPS) |
| $\Psi C_h$ | Workload needed for hosting one CNVM (GOPS) |
| $\rho_r$ | Number of active users connected to RRH node $r$ |
| n | Maximum number of physical resources blocks for cell ($r$) |
| pb | Number of physical resource blocks per user |
| $\lambda R_r$ | RRH node $r$ traffic demand (Gbps); calculated as: $[(pb/n) \cdot cp \cdot \rho_r]$, where $r \in R$ |
| $\nabla_{p,q}$ | Intra-traffic between core network VMs (CNVM) at hosting nodes $p$, and $q$ (Gbps) |
| $\alpha$ | The ratio of the backhaul to the fronthaul traffic (unitless) |
| $\Omega U$ | ONU maximum power consumption (W) |
| $\Omega L$ | OLT maximum power consumption (W) |
| $\Omega Ld$ | OLT idle power (W) |
| $CL$ | OLT maximum capacity (Gbps) |
| $CU$ | ONU maximum capacity (Gbps) |
| $\Omega R_x$ | Power consumption of the Remote Radio Head (RRH) connected to ONU node x (W) |
| $\Omega S$ | Server maximum power consumption (W) |
| $\Omega Sd$ | Server idle power (W) |
| $\Omega H_h$ | Maximum power consumption of hosting VMs at node $h$ |
| $\beta$ | Large number (unitless) |
| $\eta$ | Very small number (unitless) |
| $B$ | Capacity of the wavelength channel (Gbps) |
| $w$ | Number of wavelengths per fiber |
| $\Omega T$ | Transponder power consumption (W) |
| $\Omega RP$ | Router power consumption per port (W) |
| $\Omega G$ | Regenerator power consumption (W) |
| $\Omega E$ | EDFA power consumption (W) |
| $NG_{m,n}$ | Number of regenerators in the optical link $(m, n)$ |
| $S$ | Maximum span distance between EDFAs (km) |
| $D_{m,n}$ | Distance between node pair $(m, n)$ in the IP over WDM network (km) |
| $A_{m,n}$ | Number of EDFAs between node pair $(m, n)$ calculated as $A_{m,n} = ((D_{mn}/S) - 1) + 2$ |

TABLE III
ENERGY-EFFICIENT NFV MILP MODEL VARIABLES

| Variables | Comments |
|---|---|
| $\lambda B_{p,h}$ | Traffic from CNVMs in node $p$ to the BBUVMs in node $h$ (Gbps) |
| $\lambda R_{h,r}$ | Traffic from BBUVMs in node $h$ to the RRH node $r$ (Gbps) |
| $\sigma B_{h,r}$ | Binary indicator, set to 1 if the node $h$ hosts BBUVMs to serve the RRH node $r$, 0 otherwise |
| $\sigma B_h$ | Binary indicator, set to 1 if the node $h$ hosts a BBUVM, 0 otherwise |
| $\sigma E_{p,h}$ | Binary indicator, set to 1 if the node $h$ hosts CNVMs to serve the BBUVMs at hosting node $h$, 0 otherwise |
| $\sigma E_p$ | Binary indicator, set to 1 if the hosting node $p$ hosts CNVMs is, 0 otherwise |
| $\psi_{p,q}$ | Binary indicator, set to 1 if two different hosting nodes $p$ and $q$ host CNVMs, 0 otherwise. It is equivalent to the ANDing of the two binary variables ($\sigma E_p$, $\sigma E_q$). |
| $\sigma \chi_h$ | Binary indicator, set to 1 if the hosting node $h$ hosts any virtual machine of any type, 0 otherwise. It is equivalent to the ORing of the two binary variables ($\sigma B_h$, $\sigma E_h$). |
| $\lambda E_{p,q}$ | Traffic between hosting nodes due to CNVMs communication (Gbps) |
| $\lambda T_{p,q}$ | Total traffic from node $p$ to node $q$ caused by CNVM to CNVM traffic and CNVM to BBUVM traffic (Gbps) |
| $\lambda R_{x,y}^{h,r}$ | Traffic from hosting node $h$ to RRH node $r$ that traverses the link between the nodes $(x, y)$ in the network in Gb/s |
| $\lambda T_{x,y}^{p,q}$ | Total traffic from node $p$ to node $q$ that traverses the link between the nodes $(x, y)$ in the network (Gbps) |
| $\Psi B_h$ | BBU workload at node $h$ (GOPS) |
| $\Psi i_h$ | The integer part of the total normalized workload at node |
| $\Psi f_h$ | The fractional part of the total normalized workload at node $h$. |
| $W_{i,j}$ | Number of wavelength channels in the virtual link $(i,j)$ |
| $W_{m,n}^{i,j}$ | Number of wavelength channels in the virtual link $(i,j)$ that traverse the physical link $(m,n)$ |
| $f_{m,n}$ | Number of fibers in the physical link $(m,n)$ |
| $W_{m,n}$ | Total number of wavelengths in the physical link $(m,n)$ |
| $\Lambda_m$ | Number of aggregation ports of the router at node $m$ |

The total power consumption is composed of:
1) The power consumption of RRHs and ONUs

$$\sum_{x \in U} \left[ \Omega R_x + \frac{\Omega U}{CU} \cdot \left( \sum_{h \in H} \sum_{r \in R} \sum_{y \in TN_x} \lambda R_{x,y}^{h,r} + \sum_{p \in H} \sum_{q \in H: p \neq q} \sum_{y \in TN_x \cap H} \lambda T_{x,y}^{p,q} \right) \right]$$

2) The power consumption of the OLTs

$$\sum_{x \in L} \left[ \Omega Ld + \frac{\Omega L - \Omega Ld}{CL} \cdot \left( \sum_{h \in H} \sum_{r \in R} \sum_{y \in TN_x} \lambda R_{x,y}^{h,r} + \sum_{p \in H} \sum_{q \in H: p \neq q} \sum_{y \in TN_x \cap H} \lambda T_{x,y}^{p,q} \right) \right]$$

3) The power consumption of the IP over WDM network

$$\left( \Omega RP \cdot \sum_{m \in N} \Lambda_m \right) + \left( \Omega RP \cdot \sum_{m \in N} \sum_{n \in NN_m} W_{m,n} \right)$$
$$+ \left( \Omega T \cdot \sum_{m \in N} \sum_{n \in NN_m} W_{m,n} \right) + \left( \Omega E \cdot \sum_{m \in N} \sum_{n \in NN_m} A_{m,n} \cdot f_{m,n} \right)$$
$$+ \left( \Omega G \cdot \sum_{m \in N} \sum_{n \in NN_m} NG_{m,n} \cdot W_{m,n} \right)$$

4) The total power consumption of VMs and hosting servers

$$\sum_{h \in H} \left( \Omega Sd \cdot (\Psi i_h + \sigma \chi_h) + \Psi f_h \cdot (\Omega S - \Omega Sd) \right)$$

The model objective is to minimize the total power consumption as follows:

Minimize

$$\sum_{x \in U} \left[ \Omega R_x + \frac{\Omega U}{CU} \cdot \left( \sum_{h \in H} \sum_{r \in R} \sum_{y \in TN_x} \lambda R_{x,y}^{h,r} + \sum_{p \in H} \sum_{q \in H: p \neq q} \sum_{y \in TN_x \cap H} \lambda T_{x,y}^{p,q} \right) \right]$$

$$+ \sum_{x \in L} \left[ \Omega Ld + \frac{\Omega L - \Omega Ld}{CL} \cdot \left( \sum_{h \in H} \sum_{r \in R} \sum_{y \in TN_x} \lambda R_{x,y}^{h,r} + \sum_{p \in H} \sum_{q \in H: p \neq q} \sum_{y \in TN_x \cap H} \lambda T_{x,y}^{p,q} \right) \right]$$

$$+ \left( \Omega RP \cdot \sum_{m \in N} \Lambda_m \right) + \left( \Omega RP \cdot \sum_{m \in N} \sum_{n \in NN_m} W_{m,n} \right)$$

$$+ \left( \Omega T \cdot \sum_{m \in N} \sum_{n \in NN_m} W_{m,n} \right) + \left( \Omega E \cdot \sum_{m \in N} \sum_{n \in NN_m} A_{m,n} \cdot f_{m,n} \right)$$

$$+ \left( \Omega G \cdot \sum_{m \in N} \sum_{n \in NN_m} NG_{m,n} \cdot W_{m,n} \right)$$

$$+ \sum_{h \in H} \left( \Omega Sd \cdot (\Psi i_h + \sigma \chi_h) + \Psi f_h \cdot (\Omega S - \Omega Sd) \right)$$

Subject to the following constraints:
1) Traffic from CNVM to BBUVM

$$\sum_{p \in H} \lambda B_{p,h} = \alpha \cdot \sum_{r \in R} \lambda R_{h,r} \quad (12)$$
$$\forall h \in H$$

2) Traffic to RRH nodes

$$\sum_{h \in H} \lambda R_{h,r} = \lambda R_r \quad (13)$$
$$\forall r \in R.$$

Constraint (12) represents the traffic from CNVMs to the BBUVM in node $h$ where $\alpha$ is a unitless quantity which represents the ratio of backhaul to fronthaul traffic. Note that this constraint allows a BBUVM to receive traffic from more than a single CNVM, which may occur for example in network slicing.
Constraint (13) represents the traffic to RRH nodes from all BBUVMs that are hosted in hosting nodes. This enables an RRH to receive traffic from more than a single BBUVM (network slicing).

3) The served RRH nodes and the location of BBUVM

$$\beta \cdot \lambda R_{h,r} \geq \sigma B_{h,r} \quad (14)$$
$$\forall r \in R, \forall h \in H,$$

$$\lambda R_{h,r} \leq \beta \cdot \sigma B_{h,r} \quad (15)$$
$$\forall r \in R, \forall h \in H,$$

$$\beta \cdot \sum_{\forall r \in R} \lambda R_{h,r} \geq \sigma B_h \quad (16)$$
$$\forall h \in H$$

$$\sum_{\forall r \in R} \lambda R_{h,r} \leq \beta \cdot \sigma B_h \quad (17)$$
$$\forall h \in H$$

Constraint (14) and (15) ensure that the RRH node $r$ is served by the BBUVM that is hosted at node $h$ as illustrated in Fig. 7. Constraints (16) and (17) determine the location of BBUVM; $\beta$ is a large enough number to ensure that $\sigma B_{hr}$ and $\sigma B_h$ are equal to 1 when $\sum_{\forall r \in R} \lambda R_{hr} > 0$. In constraint (16) there are two possibilities for the value of $(\sum_{\forall r \in R} \lambda R_{h,r})$ which are either zero (no traffic from $h$ to $r$) or greater than zero (there is a traffic from $h$ to $r$). When the value of $\sum_{\forall r \in R} \lambda R_{h,r}$ is zero, the left-hand side of the inequality $(\beta \cdot \sum_{\forall r \in R} \lambda R_{h,r})$ should be zero and this sets the value of $\sigma B_h$ to zero. In the second case when the value of $\sum_{\forall r \in R} \lambda R_{h,r}$ is greater than zero, the left-hand side of the inequality $(\beta \cdot \sum_{\forall r \in R} \lambda R_{h,r})$ will be much greater than 1 because of the large value $\beta$. In this, the value of $\sigma B_h$ may be set to 1 or zero. In the same way constraint (17) sets the value of $\sigma B_h$. Table V illustrates the operation of constraints (16) and (17).

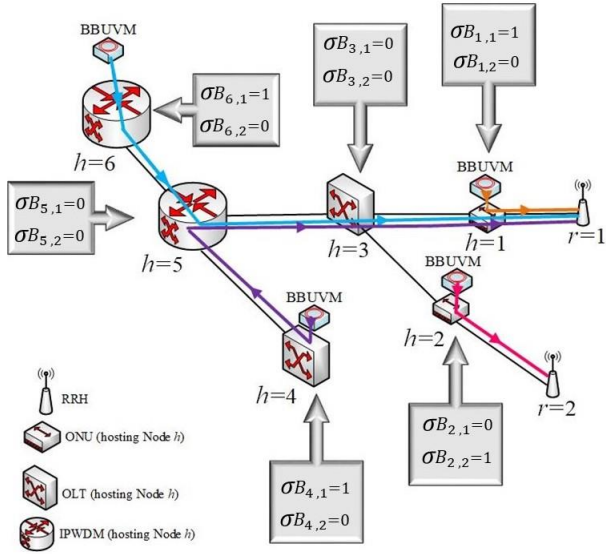

Fig. 7. BBUVM and the traffic toward RRH nodes

4) CNVM locations

$$\beta \cdot \lambda B_{p,h} \geq \sigma E_{p,h} \quad (18)$$
$$\forall p, q \in H, p \neq q$$

$$\lambda B_{p,h} \leq \beta \cdot \sigma E_{p,h} \quad (19)$$
$$\forall p, q \in H, p \neq q$$

$$\sigma E_p \geq \eta \cdot \sum_{h \in H} \lambda B_{p,h} \quad (20)$$
$$\forall p \in H$$

$$\sigma E_p \leq 1 + \sum_{h \in H} \lambda B_{ph} - \eta \quad (21)$$
$$\forall p \in H$$

$$\psi_{pq} \leq \sigma E_p \quad (22)$$
$$\forall p, q \in H, p \neq q$$

$$\psi_{p,q} \leq \sigma E_q \quad (23)$$
$$\forall p, q \in H, p \neq q$$

$$\psi_{p,q} \geq \sigma E_p + \sigma E_q - 1 \quad (24)$$
$$\forall p, q \in H, p \neq q$$

5) Hosting any VM of any type

$$\sigma \chi_h \leq \sigma B_h + \sigma E_h \quad (25)$$
$$\forall h \in H$$

$$\sigma \chi_h \geq \sigma B_h \quad (26)$$
$$\forall h \in H$$

$$\sigma \chi_h \geq \sigma E_h \quad (27)$$
$$\forall h \in H$$

TABLE V
BBUVM CONSTRAINTS OPERATION

| Input | Constraint | Outcome | $\sigma B_h$ | Value of $\sigma B_h$ that satisfies both constraints |
|---|---|---|---|---|
| $\sum_{\forall r \in R} \lambda R_{h,r} > 0$ | $\beta \cdot \sum_{\forall r \in R} \lambda R_{h,r} \geq \sigma B_h$ | $\beta \cdot \sum_{\forall r \in R} \lambda R_{h,r} \gg 1$ | 0 or 1 | 1 |
| | $\sum_{\forall r \in R} \lambda R_{h,r} \leq \beta \cdot \sigma B_h$ | $\beta \cdot \sigma B_h \gg 1$ | 1 | |
| $\sum_{\forall r \in R} \lambda R_{h,r} = 0$ | $\beta \cdot \sum_{\forall r \in R} \lambda R_{h,r} \geq \sigma B_h$ | $\beta \cdot \sum_{\forall r \in R} \lambda R_{h,r} = 0$ | 0 | 0 |
| | $\sum_{\forall r \in R} \lambda R_{h,r} \leq \beta \cdot \sigma B_h$ | $\beta \cdot \sigma B_h = 0$ | 0 or 1 | |

Constraints (18) and (19) ensure that the BBUVMs at node $h$ are served by the CNVMs that are hosted at the node $p$. Constraints (20) and (21) determine the location of the CNVMs by setting the binary variable $\sigma E_p$ to 1 if there is a CNVM hosted at node $p$, where $\eta$ is very small number. Fig. 8 illustrates the functions of constraints (20) and (21) whilst Table VI illustrates their operation. Constraints (22) - (24) ensure that the CNVMs communicate with each other if they are hosted at different nodes $p$ and $q$, and this is equivalent to the logical operation $\psi_{p,q} = \sigma E_p$ AND $\sigma E_q$. Fig. 9 illustrates the function of constraints (22) - (24). Constraints (25) - (27) determine if the hosting node $h$ hosts any VM of any type (BBUVM or CNVM). It is equivalent to the logical operation $\sigma \chi_h = \sigma E_p$ OR $\sigma E_q$

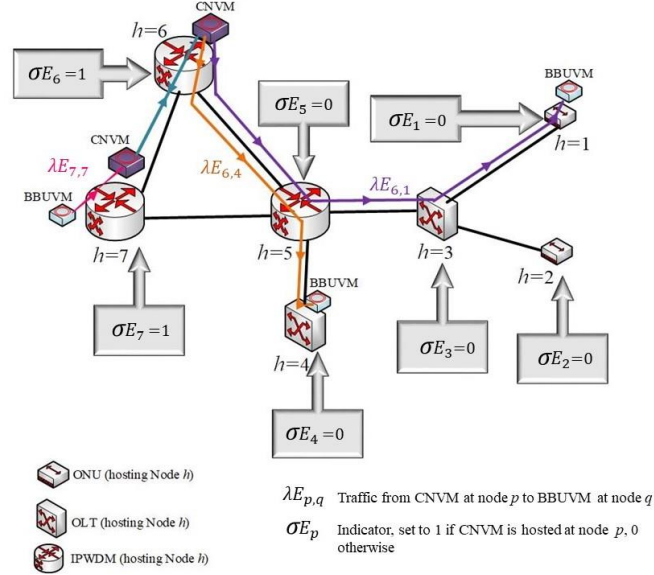

Fig. 8. CNVM and the traffic toward BBUVMs nodes

TABLE VI
CNVM CONSTRAINTS OPERATION

| Input | Constraint | Outcome | $\sigma B_h$ | Value of $\sigma E_p$ that satisfies both constraints |
|---|---|---|---|---|
| $\sum_{h \in H} \lambda B_{ph} > 0$ | $\sigma E_p \geq \eta \cdot \sum_{h \in H} \lambda B_{p,h}$ | $\eta \cdot \sum_{h \in H} \lambda B_{p,h} \ll 1$ | 1 | 1 |
| | $\sigma E_p \leq 1 + \sum_{h \in H} \lambda B_{ph} - \eta$ | $1 + \sum_{h \in H} \lambda B_{ph} - \eta > 1$ | 0 or 1 | |
| $\sum_{h \in H} \lambda B_{ph} = 0$ | $\sigma E_p \geq \eta \cdot \sum_{h \in H} \lambda B_{p,h}$ | $\eta \cdot \sum_{h \in H} \lambda B_{p,h} = 0$ | 0 or 1 | 0 |
| | $\sigma E_p \leq 1 + \sum_{h \in H} \lambda B_{ph} - \eta$ | $1 + \sum_{h \in H} \lambda B_{ph} - \eta < 1$ | 0 | |

6) Communication traffic between CNVMs

$$\lambda E_{p,q} = \nabla_{p,q} \cdot \psi_{p,q} \qquad (28)$$
$$\forall p, q \in H: p \neq q$$

7) Total traffic between two hosting nodes

$$\lambda T_{p,q} = \lambda E_{p,q} + \lambda B_{p,q} \qquad (29)$$
$$\forall p, q \in H: p \neq q$$

8) Flow conservation of the total traffic to the RRH nodes

$$\sum_{y \in TN_x} \lambda R_{x,y}^{h,r} - \sum_{y \in TN_x} \lambda R_{y,x}^{h,r} = \begin{cases} \lambda R_{h,r} & \text{if } x = h \\ -\lambda R_{h,r} & \text{if } x = r \\ 0 & \text{otherwise} \end{cases} \qquad (30)$$
$$\forall r \in R, \forall h \in H, \forall x \in T$$

9) Flow conservation of hosting nodes communication traffic

$$\sum_{y \in TN_x \cap H} \lambda T_{x,y}^{p,q} - \sum_{y \in TN_x \cap H} \lambda T_{y,x}^{p,q}$$
$$= \begin{cases} \lambda T_{p,q} & \text{if } x = p \\ -\lambda T_{p,q} & \text{if } x = q \\ 0 & \text{otherwise} \end{cases} \qquad (31)$$
$$\forall p, q, x \in H: p \neq q$$

Constraint (28) represents the traffic between CNVMs at hosting nodes $p$ and $q$. Constraint (29) represents the total traffic between any two hosting nodes $(p, q)$ which is caused by virtual machines communication. Constraint (30) represents the flow conservation of the total fronthaul traffic to the RRH nodes. Fig. 10 illustrates the principle of flow conservation, and for clarification purposes, it is applied to constraint (30). Constraint (31) represents the flow conservation of the total traffic between any two hosting nodes that might host virtual machines of any type (BBUVM or CNVM).

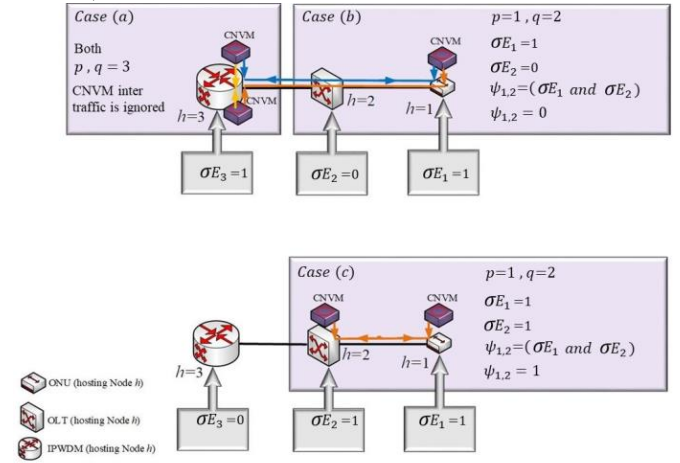

Fig. 9. CNVM and the common locus

10) Total BBU workload at any hosting node $h$

$$\Psi B_h = \left( \left( \sum_{\forall r \in R} \lambda R_{h,r} \right) / cp \cdot \right) \Psi X \qquad (32)$$
$$\forall h \in H$$

11) Total normalized workload at hosting node $h$

$$\Psi i_h + \Psi f_h = (\Psi B_h + \Psi C_h)/\Psi S \quad (33)$$
$$\forall h \in H$$

12) Hosting node capacity

$$\left(\Omega Sd \cdot (\Psi i_h + \sigma \chi_h) + \Psi f_h \cdot (\Omega S - \Omega Sd)\right) \leq \Omega H_h \quad (34)$$
$$\forall h \in H$$

13) GPON link constraints

$$\sum_{h \in H} \sum_{r \in R} \sum_{j \in TN_i \cap L} \lambda R_{i,j}^{h,r} \leq 0 \quad (35)$$
$$\forall i \in U$$

$$\sum_{p \in H} \sum_{q \in H, q \neq p} \sum_{j \in TN_i \cap L} \lambda T_{i,j}^{p,q} \leq 0 \quad (36)$$
$$\forall i \in U$$

$$\sum_{h \in H} \sum_{r \in R} \sum_{j \in TN_i \cap N} \lambda R_{i,j}^{h,r} \leq 0 \quad (37)$$
$$\forall i \in L$$

$$\sum_{p \in H} \sum_{q \in H, q \neq p} \sum_{j \in TN_i \cap N} \lambda T_{i,j}^{p,q} \leq 0 \quad (38)$$
$$\forall i \in L$$

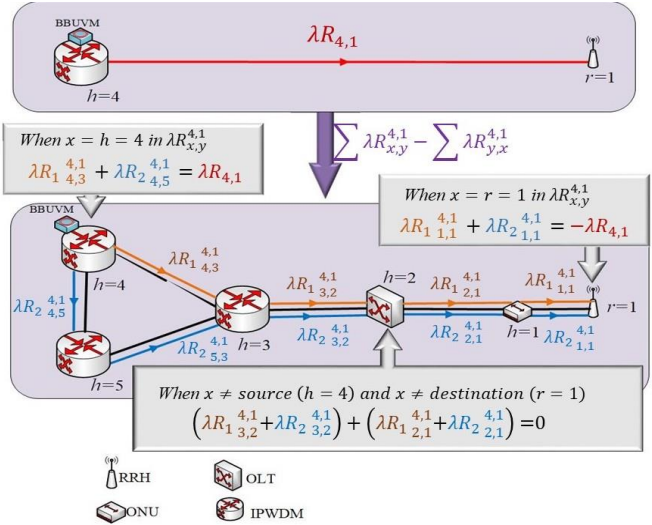

Fig. 10. Flow conservation principle

Constraint (32) represents the total BBU workload at any hosting node $h$. Constraint (33) calculates the total BBU and CNVM normalized workload at any hosting node. The workload is scaled and normalized relative to the server CPU workload and is separated into integer and fractional parts. Constraint (34) ensures that the total power consumption of hosting VMs does not exceed the maximum power consumption allocated for each host. Constraints (35) – (38) ensure that the total PON downlink traffic does not flow in the opposite direction.

14) Virtual Link capacity of the IP over WDM network

$$\sum_{p \in H} \sum_{q \in H, q \neq p} \lambda T_{i,j}^{p,q} + \sum_{h \in H} \sum_{r \in R} \lambda R_{i,j}^{h,r} \leq W_{i,j} \cdot B \quad (39)$$
$$\forall i, j \in N, i \neq j.$$

15) Flow conservation in the optical layer of IP over WDM network

$$\sum_{n \in NN_m} W_{m,n}^{i,j} - \sum_{n \in NN_m} W_{n,m}^{i,j} = \begin{cases} W_{i,j} & \text{if } n = i \\ -W_{i,j} & \text{if } n = j \\ 0 & \text{otherwise} \end{cases} \quad (40)$$
$$\forall i, j, m \in N, i \neq j$$

Constraint (39) ensures that the total traffic traversing the virtual link $(i,j)$ does not exceed its capacity, in addition it determines the number of wavelength channels that carry the traffic burden of that link. Constraint (40) represents the flow conservation in the optical layer of the IP over WDM network. It ensures that the total expected number of incoming wavelengths for the IP over WDM nodes of the virtual link $(i,j)$ is equal to the total number of outgoing wavelengths of that link.

16) Number of wavelength channels

$$\sum_{i \in N} \sum_{j \in N: i \neq j} W_{m,n}^{i,j} \leq w \cdot f_{m,n} \quad (41)$$
$$\forall m \in N, \forall n \in NN_m$$

17) Total number of wavelength channels

$$W_{m,n} = \sum_{i \in N} \sum_{j \in N: i \neq j} W_{m,n}^{i,j} \quad (42)$$
$$\forall m \in N, \forall n \in NN_m$$

18) Number of aggregation ports

$$\Lambda_i = \left( \sum_{j \in L \cap TN_i} \left( \sum_{p \in H} \sum_{q \in H, q \neq p} \lambda T_{i,j}^{p,q} + \sum_{h \in H} \sum_{r \in R} \lambda R_{i,j}^{h,r} \right) \right) / B \quad (43)$$
$$\forall i \in N$$

Constraints (41) and (42) are the constraints of the physical link $(m,n)$. Constraint (41) ensures that the total number of wavelength channels in the logical link $(i,j)$ that traverse the physical link $(m,n)$ does not exceed the fiber capacity. Constraint (42) determines the number of wavelength channels in the physical link and ensures that it is equals to the total number of wavelength channels in the virtual link traversing that physical link. Constraint (43) determines the required number of aggregation ports in each IP over WDM router.

V. MILP MODEL SETUP AND RESULTS

Five IP over WDM nodes are considered constituting the optical backbone network of the proposed architecture. The distribution and topology of the IP over WDM nodes have been built upon the NSFNET network described in [46-51]. Each IP over WDM node in turn is attached to two GPONs with one OLT and two ONUs for each GPON. Accordingly, the network topology has 10 OLTs and 20 ONUs. In addition, each ONU is connected to one RRH node as shown in Fig. 11. Two GPONs for each IP over WDM node are enough to investigate the VM response for demands and power savings. To finalize the portrait of the network topology, we have concentrated on the distribution of the hosting nodes and the way in which they are connected to each other and for this reason the GPON splitters are not shown.

As alluded to earlier, two types of VMs have been considered: BBUVM, which realize the functions of the BBU, and CNVM to achieve the functions of the mobile core network. The amount of workload needed for BBUVMs is calculated in GOPS according to (11) [45] and based on the calculated workload, the hosting server CPU utilization due to hosting BBUVMs is determined. On the other hand, the total workload needed for CNVMs is calculated based on the number BBUVMs group in each hosting node since we have allocated one CNVM for each group of BBUVMs in one hosting node. A single VM consumes around 18W [52] and by knowing the hosting server maximum power consumption (365W), idle power (112) and the maximum workload (368 GOPS), $\Psi C_h$ can calculated for a single VM. Therefore $\Psi C_h =$ corresponds is $(18 \times 368)/(365 - 112) = 26$ GOPS.

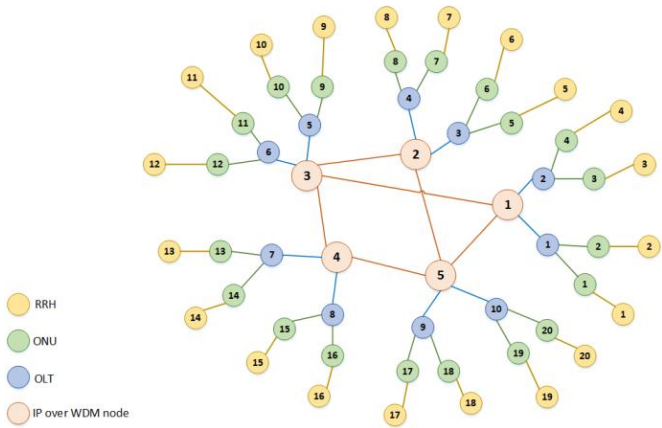

Fig. 11.  Tested network topology

We have investigated the effect of the intra-traffic between the CNVMs by considering a range of intra-traffic relative to the total network traffic (0%, 1%, 5%,10%, and 16% of the total traffic) flows from CNVMs. Moving toward the access network, each RRH node is considered to serve a small cell that operates on 10 MHz bandwidth and with a maximum capacity of 10 users. Each user in the small cell is allocated 5 physical resources blocks (PRB) as the users are assumed to request the same task from the network. Accordingly, the total downlink traffic to the RRH node depends on the total number of active users in the small cell. The input parameters to the developed MILP model are listed in Table VII. We have considered 17 time slots over all the day from 0 hour to 24 hour in steps of 1.5 hours using the average number of users daily profile shown in Fig. 12. The MILP results are compared with the case where there is no NFV deployment. In the "no virtualization" scenario, the BBU is located close to the RRH where they are attached to each other, whilst the integrated platform ASR5000 is deployed to realize mobile core network functionalities and it is connected directly to the IP over WDM network. The ASR5000 maximum power consumption, idle power, and maximum capacity are 5760 (W), 800 (W), and 320 (Gbps) respectively [53], whilst the BBU maximum power consumption, idle power, and maximum capacity are 531 (W), 51 (W), 9.8 (Gbps) respectively [54].

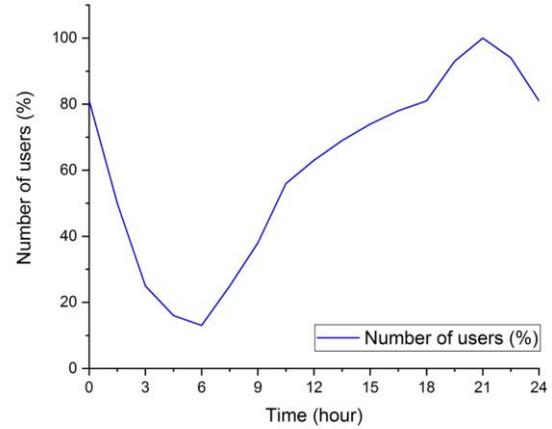

Fig. 12.  Average number of users daily profile [1]

TABLE VII
MILP MODEL INPUT PARAMETERS

| Parameters | Comments |
| --- | --- |
| Line coding rate for 8B/10B line coding ($l$) | 10/8 (bit / sample) |
| Number of MIMO layers ($y$) | 2 |
| Number of bits used in QAM modulation for 64 QAM modulation ($q$) | 6 (bits) |
| Number of antennas in a cell ($a$) | 2 |
| Maximum fronthaul (CPRI) data rate for CPRI line rate option 7 ($cp$) | 9.8304 (Gbps) [42] |
| Maximum baseband processing workload needed for fully loaded RRH ($\Psi X$) given by: $30 \cdot a + 10 \cdot a^2 + 20 \cdot q \cdot l \cdot y$ | 400 (GOPS) |
| Server CPU maximum workload ($\Psi S$) | 368 (GOPS) [55] |
| Workload needed for hosting one CNVM ($\Psi C$) | 26.17 (GOPS) |
| Number of active users in a small cell ($\rho_r$) | Uniformly distributed (1-10 users) |
| Maximum number of users per cell ($n$) | 10 (users) |
| Number of physical resource blocks per user ($pb$) | 5 (PRB) |
| The ratio of the backhaul to the front haul traffic ($\alpha$) | 0.1344 (unitless) |
| ONU maximum power consumption ($\Omega U$) | 15 (W) [56] |
| OLT maximum power consumption ($\Omega L$) | 1940 (W) [57] |
| OLT idle power ($\Omega Ld$) | 60 (W) [57] |
| OLT maximum capacity ($CL$) | 8600 (Gbps) [57] |
| ONU maximum capacity ($CU$) | 10 (Gbps) [56] |
| RRH node power consumption ($\Omega R_x$) | 1140 (W) [58] |
| Hosting server maximum power consumption ($\Omega S$) | 365 (W) [59] |
| Hosting server idle power consumption ($\Omega Sd$) | 112 (W) [59] |
| Capacity IP over WDM wavelength channel ($B$) | 40 (Gbps) [60-62] |
| Number of wavelengths per fiber in IP over WDM ($w$) | 32 [60] |
| Transponder power consumption ($\Omega T$) | 167 (W) [63] |
| Router port power consumption ($\Omega RP$) | 825 (W) [64] |
| Regenerator power consumption ($\Omega G$) | 334 (W) [64] |
| EDFA power consumption ($\Omega E$) | 55 (W) [64] |
| Maximum span distance between EDFAs ($S$) | 80 (km) [60, 61] |

The results in Fig. 13 show the total power consumption of the of the case where no virtualization is deployed (standard model) as well as the cases where the virtualization is deployed under different CNVMs intra-traffic for different time slots in a day. Fig. 14 shows the total power consumption of the same scenarios versus the total number of active users in the networks. The virtualization model has resulted in less power consumption compared to the no virtualization model (standard model) as it optimizes the processing locations of the downlink traffic through optimum placement and consolidation of VMs.

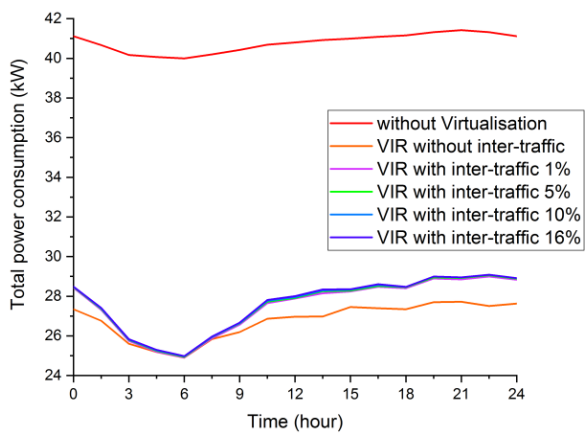

Fig. 13. Total power consumption without and with virtualization under different CNVMs inter-traffic at different time slots of a day

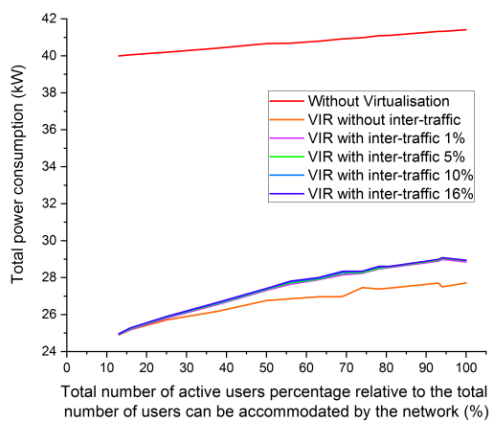

Fig. 14. Total power consumption without and with virtualization under different CNVMs inter-traffic versus total active users in the network

Fig. 15 compares the total power saving of the virtualization model under different CNVMs inter-traffic for one day while Fig. 16 show the virtualization power saving under different CNVMs inter-traffic versus total number of active users. Compared to other virtualization cases, virtualization without CNVMs inter-traffic has saved a maximum of 38% (average 34%). This is because there is no power consumed by the CNVMs inter-traffic as this traffic is zero. The total power saving decreases as the CNVMs inter-traffic increases to reach its lowest value in the case of virtualization with 16% CNVMs inter-traffic which is 37% (average 32%).

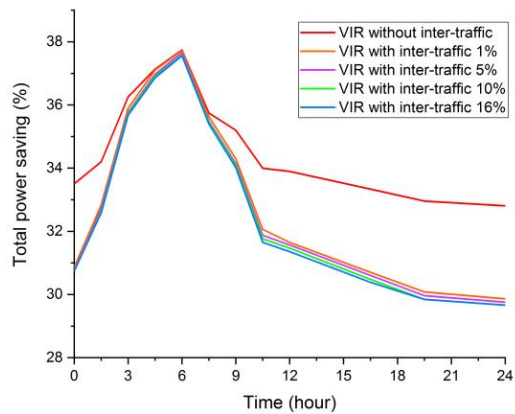

Fig. 15. Power saving comparison of virtualization under different CNVMs inter-traffic for a day

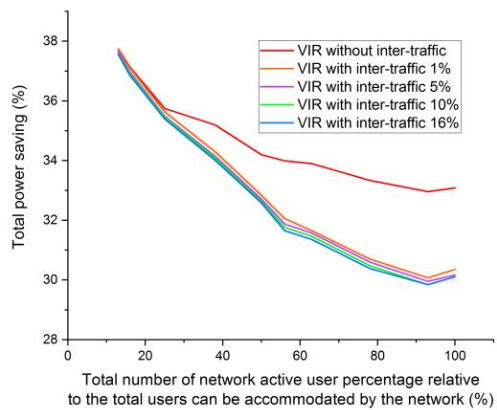

Fig. 16. Power saving comparison of virtualization under different CNVMs inter-traffic versus total number of active users

Virtualization in the presence of CNVMs inter-traffic resulted in comparable values of total power consumption (and power saving) for all values of CNVMs inter-traffic greater than zero. The main reason behind this is that the CNVMs inter-traffic produces relatively small amount of power consumption compared to the power consumption induced by the fronthaul traffic and hosting server as shown in Fig. 17. As the inter-traffic increases, the MILP model tends to eliminate its effect by consolidating CNVMs in one place.

Although virtualization has saved a maximum of 38% (without CNVMs inter-traffic) and 37% (with 16% CNVMs inter-traffic) of the total power consumption, it cannot provide such level of power saving over the entire day. As the number of active users varies with the time of day (as in Fig. 12), the power saving achieved by virtualization varies accordingly. The results in Figs. 15 and 16 show that a high-power saving is achieved when the total number of active users is around 20% (around 4 am to 8 am) while the lowest power saving is recorded at high number of active users (during the day rush hours). At small number of active users, the MILP model tends to consolidate all the VMs in the IP over WDM network to minimize the number of servers hosting VMs to reduce the total power consumption.

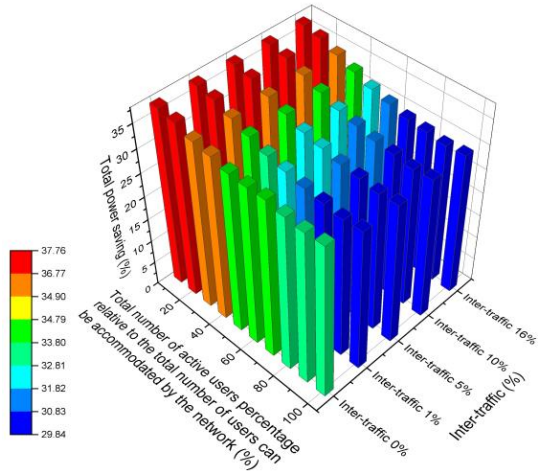

Fig. 17. 3-Dimensional presentation of the total power saving for virtualization under different CNVMs inter-traffic versus different number of active users in the network.

Figs. 18 and 19 show the VMs consolidation and distribution over the network at low number of active users (13%) under CNVMs of 0% and 16% respectively. At low number of active users and 0% inter-traffic, the MILP model consolidates the VMs at the IP over WDM network. Since the total number of active users is low, the fronthaul traffic is relatively low and consequently the power consumption induced by the fronthaul traffic is low compared to the hosting power consumption (servers power). For this reason, the MILP model tends to pack BBUVMs in the IP over WDM network as much as possible to reduce the power consumed by the hosting servers. Also, the MILP model tends to host CNVMs close to the BBUVMs as the inter-traffic between CNVMs is zero. Once the inter-traffic is greater than zero, the MILP model consolidates the CNVMs at one location as in Fig. 19.

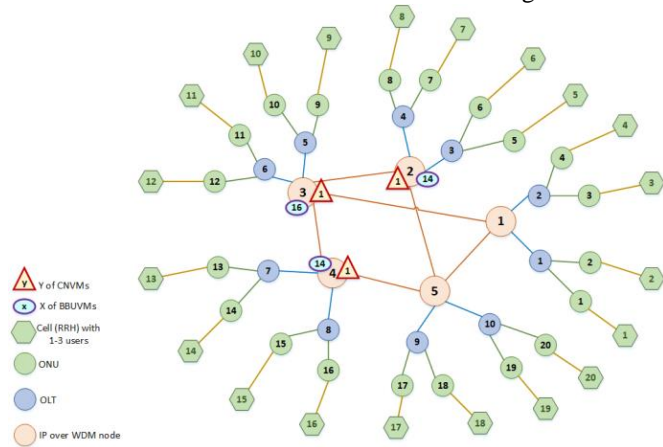

Fig. 18. VMs distribution over network under active users 13% of the total network capacity without CNVMs inter-traffic.

Figs. 20 and 21 show the VMs consolidation and distribution over the network with high number of active users (around 100%) under 0% and 16% CNVMs inter-traffic. When the number of active users is high, the amount of fronthaul traffic is high, for that reason the MILP model tends to distribute the BBUVMs at the closest centralized location to the users which are the OLTs, while CNVMs inter-traffic has no effect on the distribution of BBUVMs.

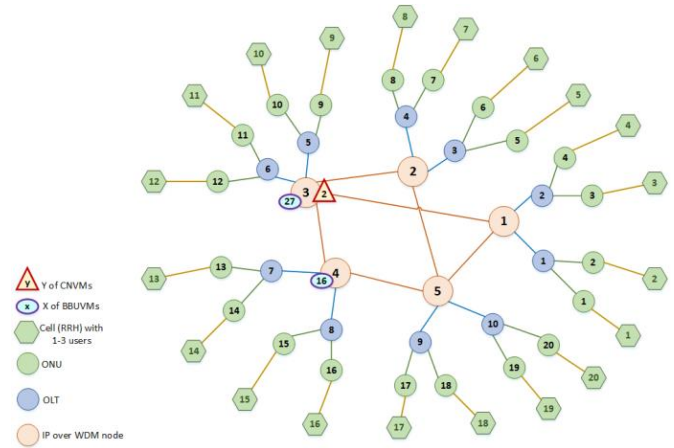

Fig. 19. VMs distribution over network under active users 13% of the total network capacity and 16% CNVMs inter-traffic

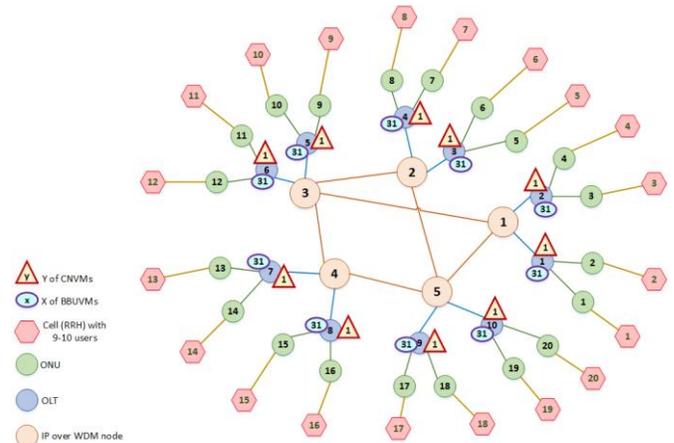

Fig. 20. VMs distribution over network under active users 100% of the total network capacity without CNVMs inter-traffic

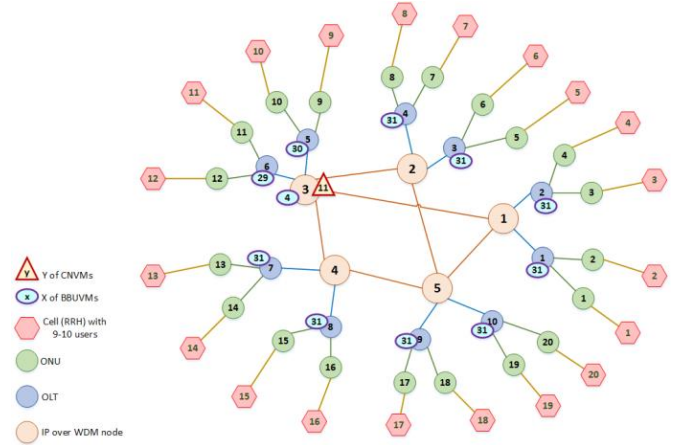

Fig. 21. VMs distribution over network under active users 100% of the total network capacity and 16% CNVMs inter-traffic.

Hosting BBUVMs in OLTs when the number of users is high ensures shorter paths for the fronthaul traffic than hosting BBUVMs in the IP over WDM networks and consequently, the power consumed by this traffic is less. For CNVMs, the MILP model tends to distribute them close to the BBUVM when there is no inter-traffic between them, and this is clearly seen in Fig. 20. In contrast, when the inter-traffic between CNVMs is greater than zero, the MILP model tends to centralize the location of CNVMS in the IP over WDM network to reduce the power consumption induced by the inter-traffic and the power of the hosting servers as shown in Fig. 21.

## VI. REAL-TIME HEURISTICS IMPLEMENTATION AND RESULTS

### A. Energy Efficient NFV with no CNVMs inter-traffic (EENFVnoITr) heuristic

The EENFVnoITr provides real-time implementation of the MILP model without CNVMs inter-traffic. The pseudocode of the heuristic is shown in Fig. 22. the network is modelled by sets of network elements *NE*, and links *L*. The heuristic obtains the network topology $G=(NE, L)$ and the physical topology of the IP over WDM network $G_p=(N, L_p)$ where $N$ is the set of IP over WDM nodes and $L_p$ is the set of physical links. The total download request (fronthaul traffic) of each RRH node is calculated based on the total number of active users in each cell (RRH). The heuristic determines the amount of baseband workload needed to process each RRH download request. According to the baseband workload for each requested download traffic and the available capacity of the hosting VM server, the EENFVnoITr heuristic chooses the closest place to accommodate BBUVM in such a way that it serves as many RRH requests as possible. The EENFVnoITr heuristic may host a BBUVM in an OLT node if it has enough processing capacity to serve all the requests from the closest RRH nodes. In this way, the heuristic exploits bin packing techniques to reduce the processing power consumption. The amount of fronthaul traffic delivered by each BBUVM determines the backhaul traffic flows from each CNVMs toward BBUVMs. The EENFVnoITr heuristic determines the total amount of backhaul traffic that may flow from each IP over WDM node and sorts them in a descending order. The nodes in the top of the sorted list of IP over WDM nodes represent highly recommended nodes to host CNVMs. In such a scenario, the EENFVnoITr heuristic ensures less of the backhaul traffic flows in the IP over WDM network. The EENFVnoITr heuristic uses the sorted list to accommodate CNVMs. Once the VMs are distributed and the logical traffic is routed, the EENFVnoITr heuristic obtains the physical graph $G_p=(N, L_p)$ and determines the traffic in each network segment. The IP over WDM network configuration such as the number of fibers, router ports, and the number of EDFA is determined the total power consumption is evaluated. The heuristic reduces the number of CNVMs candidate locations by one, re-configures the IP over WDM network, and re-evaluates the power consumption to determine the best number and location of CNVMs for minimum power consumption.

---

Energy Efficient NFV without CNVMs inter-traffic (EENFVnoITr) Heuristic
**INPUT:** $G=(NE, L)$, $G_p=(N, L_p)$
**OUTPUT:** VMs location, workloads, and distribution

1: $\forall r \in RRH$ determine number of users and calculate node demand $(r, D_r)$; where $D_r \in D$ /*$D$ is the total demands*/
2: $\forall D_r \in D$ determine BBUVM workload $\Psi_r$
3: $\forall r \in RRH$ find
    $(r, h) = min\ (shortestPath(r, \{h \in NE \cap OLT\})$
4: **if** total workload of $h >> \Psi_r$
5:     host BBUVM in $h$
6:     update workload of $h$
7:     $D_r \in D_{served}$
8: **end if**
9: $\forall D_r \notin D_{served}$ find $(r, h) = min(shortestPath(r, \{h \in N\}))$
    /*where $N$ is the IP over WDM nodes */
10:    host BBUVM in $h$
11:    update workload of $h$
12:    $D_r \in D_{served}$
13: Route the fronthaul traffic from BBUVMs to RRH nodes
14: $N' \leftarrow DESCEND\_SORT(N)$ and set $i = 1$
15: Host CNVM in $N'(i), N'(i-1), \dots N'(1)$
16: $\forall CNVM\ in\ n \in N'$ and $\forall BBUVM\ in\ h \in NE$
    find $(n, h) = min(shortestPath(n, h))$
17: Route the traffic from CNVMs to BBUVMs
18: Determine the IP over WDM network configuration
19: Determine the total power consumption (TPC)
20: **if** TPC *not* min
21:    $i = i + 1$
22:    goto 15
23: **else**
24:    EXIT
25: **end if**

Fig. 22. EENFVnoITr Pseudocode

### B. Energy Efficient NFV with CNVMs inter-traffic (EENFVwithITr) Heuristic

This section describes the energy efficient NFV with CNVMs inter-traffic heuristic (EENFVwithITr). The EENFVwithITr heuristic extends the EENFVnoITr heuristic to provide real-time implementation of the MILP model where the CNVMs are considered. The pseudocode of the heuristic is shown in Fig. 23. It uses the same approach used by EENFVnoITr, but it evaluates the CNVMs inter-traffic after the locations of CNVMs are determined.

---

Energy Efficient NFV with CNVMs inter-traffic (EENFVwithITr) Heuristic

**INPUT:** $G =(NE, L)$, $G_p = (N, L_p)$

**OUTPUT:** VMs location, workloads, and distribution

1: $\forall r \in RRH$ determine number of users and calculate node demand $(r, D_r)$; where $D_r \in D$ /*$D$ is the total demands*/
2: $\forall D_r \in D$ determine BBUVM workload $\Psi_r$
3: $\forall r \in RRH$ find
   $(r, h) = \min(shortestPath(r, \{h \in NE \cap OLT\})$
4: **if** total workload of $h \gg \Psi_r$
5:    host BBUVM in $h$
6:    update workload of $h$
7:    $D_r \in D_{served}$
8: **end if**
9: $\forall D_r \notin D_{served}$ find $(r, h) = \min(shortestPath(r, \{h \in N\}))$
   /*where $N$ is the IP over WDM nodes */
10:    host BBUVM in $h$
11:    update workload of $h$
12:    $D_r \in D_{served}$
13: Route the fronthaul traffic from BBUVMs to RRH nodes
14: $N' \leftarrow DESCEND\_SORT(N)$ and set $i = 1$
15: Host CNVM in $N'(i), N'(i-1), \dots N'(1)$
16: $\forall$ CNVM in $n \in N'$ and $\forall$BBUVM in $h \in NE$
    find $(n, h) = \min(shortestPath(n, h))$
17: Route the traffic from CNVMs to BBUVMs
18: $\forall$ CNVM in $n_x, n_y \in N'; x \neq y$
    find $(n_x, n_y) = \min(shortestPath(n_x, n_y))$
19: Route the traffic between CNVMs
20: Determine the IP over WDM network configuration
21: Determine the total power consumption (TPC)
22: **if** TPC *not* min
23:    $i = i + 1$
24:    goto 15
25: **else**
26:    EXIT
27: **end if**

Fig. 23. EENFVwithITr Pseudocode

## C. EENFVnoITr and EENFVwithITr heuristics results

In order to verify the results of the proposed MILP model, the network topology in Fig. 11 used for the MILP model is also used to evaluate the heuristics. All the parameters considered in the MILP model such as the wireless bandwidth, number of resources blocks per user, and the parameters in Table VII are considered in the evaluation of both EENFVnoITr and EENFVwithITr heuristics. The number of users allocated to each cell in the heuristics is the same as in the MILP model to ensure the requested traffic by each RRH node is the same in all models. Fig. 24 compares the total power consumption of MILP with EENFVnoITr model at different times of the day when the CNVMs inter-traffic is not considered. It is clearly seen that there is a small difference in the total power consumption of the two models and it varies over the day according to the total number of active users. The total power consumption of the MILP model is less than the EENFVnoITr heuristic with a maximum of 9% (average 5%) drop in the total power consumption. This is mainly caused by the distribution of CNVMs in the EENFVnoITr heuristic. As there is no traffic flowing between CNVMs, the EENFVnoITr accommodates them close to the BBUVMs wherever the VM servers have enough capacity. To accommodate the CNVMs, the heuristic sequentially examines the capacity of the VM servers in the OLT nodes that are close to the BBUVMs before investigating other servers in the IP over WDM network. As the distance and capacity requirements of the VM servers are met, the heuristic accommodates a CNVM in the server. This case results in high EENFVnoITr VM server power consumption compared with MILP model. This is clearly seen in Fig. 25 where the VM servers power consumption of MILP and the EENFVnoITr heuristic are compared. The total network power consumption of both EENFVnoITr heuristic and MILP model are the same for most of the time of the day. Fig. 26 shows the network power consumption of MILP model compared with EENFVnoITr heuristic.

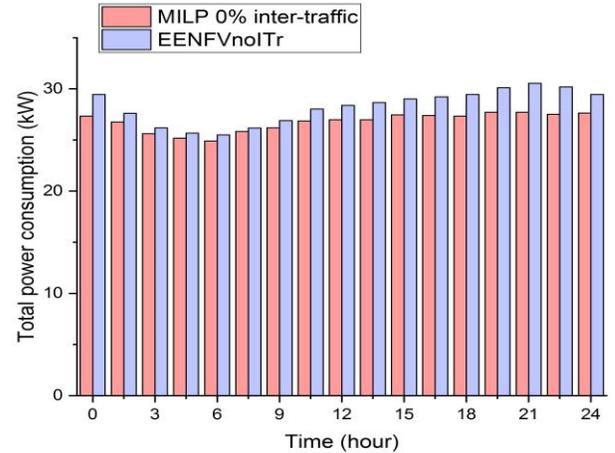

Fig. 24. Total power consumption of MILP with without CNVMS inter-traffic compared with EENFVnoITr heuristic model

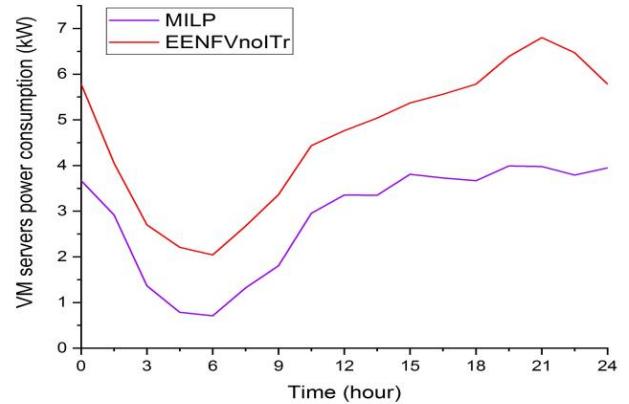

Fig. 25 VM servers power consumption of MILP model compared with EENFVnoITr heuristic

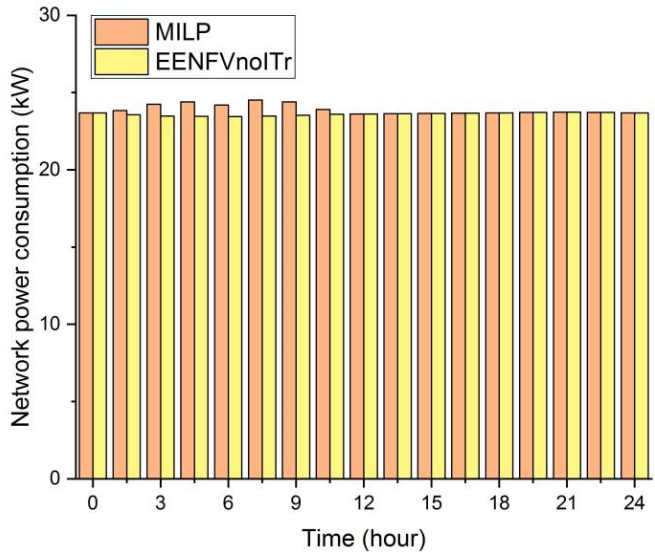

Fig. 26  Network power consumption of MILP model compared with EENFVnoITr heuristic

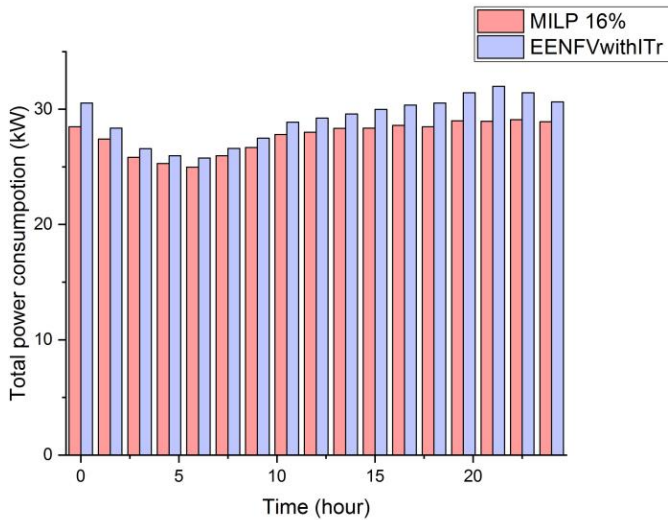

Fig. 27  Total power consumption of MILP model compared with EENFVwithITr heuristic at CNVMs inter-traffic 16% of the total backhaul traffic

It shows that there is a small difference in the network power consumption between the two models during the time of the day when the total number of active users is low. This is driven by the approach of the MILP model where it tends to accommodate the CNVMs at the IP over WDM nodes rather than OLT at the time of the day where the total number of users is low. In contrast, the heuristic tends to accommodate the CNVMs wherever the VM server is close to the BBUVMs and it has enough capacity. Fig. 27 compares the total power consumption of EENFVwithITr with the MILP model when the CNVMs inter-traffic is 16% of the total backhaul traffic. It is clearly seen that there is a small difference in the total power consumption of the two models and this varies over the day according to the total number of active users. The total power consumption of the MILP model is less than the EENFVnoITr model with a maximum drop of 9.5% (average 5%) in the total power consumption. This is mainly driven by the distribution of both CNVMs and BBUVM over the network nodes. The MILP model tends to accommodate BBUVMs and CNVMs at the IP over WDM network during times of the day when there is a small number of active users.

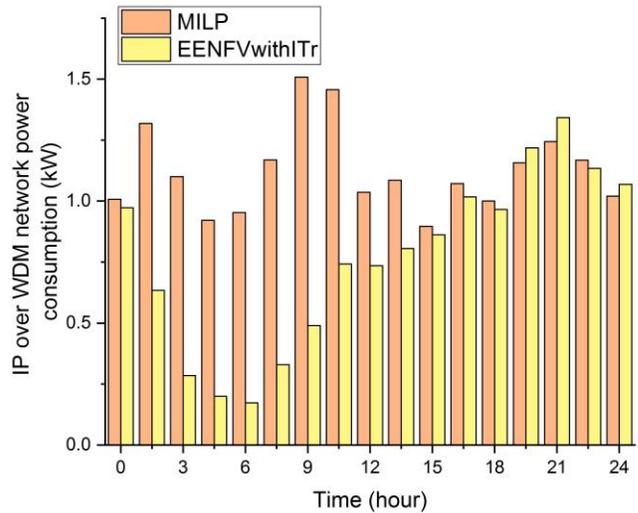

Fig. 28  IP over WDM network power consumption of MILP model compared with EENFVwithITr heuristic at CNVMs inter-traffic 16% of the total backhaul traffic

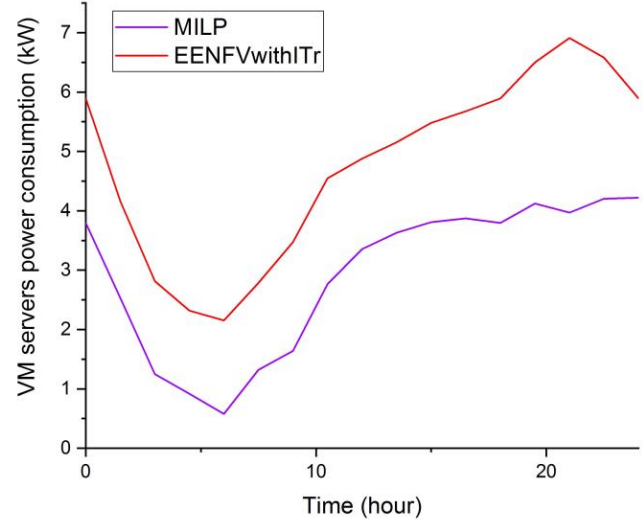

Fig. 29  VM servers power consumption of MILP model compared with EENFVwithITr heuristic at CNVMs inter-traffic 16% of the total backhaul traffic

This causes more traffic from BBUVMs and CNVMs to flow in the IP over WDM network which eventually increases the IP over WDM network power consumption as shown in Fig. 28 which compares the IP over WDM network power consumption of both MILP model and EENFVwithITr heuristic when CNVMs inter-traffic is considered 16% of the total backhaul traffic. In contrast, the IP over WDM network power consumption of EENFVwithITr varies according to the total number of active users during the day. The sequential examination by EENFVwithITr of VM servers, their location, and available capacity increases the processing distribution of VMs in the network which leads to a high VM servers power consumption compared with the MILP model as shown in Fig. 29 which compares the VM servers power consumption of the MILP model with EENFVwithITr heuristic during different times of the day.

## VII. CONCLUSIONS

This paper has investigated network function virtualization in 5G mobile networks with the impact of total number of active users in the network, the backhaul / fronthaul configurations, and the inter-traffic between VMs. A MILP optimization model was developed with the objective of minimizing the total power consumption by optimizing the VMs locations and VM servers' utilization. The MILP model results have been investigated under the impact of CNVMs traffic variation, and variation in the total number of active users during different times of the day. The MILP model results show that virtualization can save up to 38% (average 34%) of the total power consumption, also the results reveal how the total number of active users affects the BBUVMs distribution while CNVMs distribution is affected mainly by the inter-traffic between them. For real-time implementation, this paper has introduced two heuristics: Energy Efficient NFV without CNVMs inter-traffic and Energy Efficient NFV with CNVMs inter-traffic. The results obtained through the use of the heuristics were compared with the MILP model results. The comparisons showed that the total power consumption when the heuristics are used is higher than the total power consumption when the MILP optimization model is used by a maximum of 9% (average 5%).

ACKNOWLEDGMENTS

We would like to acknowledge funding from the Engineering and Physical Sciences Research Council (EPSRC) for the INTERNET (EP/H040536/1) and STAR (EP/K016873/1) projects. All data are provided in full in the results section of this paper. Mr. Ahmed Al-Quzweeni would like to acknowledge HCED for funding his PhD.

## VIII. REFERENCES

[1] G. Auer, V. Giannini, C. Desset, I. Godor, P. Skillermark, M. Olsson, *et al.*, "How much energy is needed to run a wireless network?," *IEEE Wireless Communications,* vol. 18, pp. 40-49, 2011.
[2] Cisco, "Cisco Visual Networking Index: Forecast and Methodology, 2016–2021," *White Paper,* June 2017 June 2017.
[3] I. Neokosmidis, T. Rokkas, P. Paglierani, C. Meani, K. M. Nasr, K. Moessner, *et al.*, "Techno Economic Assessment of Immersive Video Services in 5G Converged Optical/Wireless Networks," in *2018 Optical Fiber Communications Conference and Exposition (OFC)*, 2018, pp. 1-3.
[4] V. G. Nguyen, A. Brunstrom, K. J. Grinnemo, and J. Taheri, "5G Mobile Networks: Requirements, Enabling Technologies, and Research Activities," *A Comprehensive Guide to 5G Security,* pp. 31-57, 2018.
[5] Z. Zhang, Y. Gao, Y. Liu, and Z. Li, "Performance evaluation of shortened transmission time interval in LTE networks," in *Wireless Communications and Networking Conference (WCNC), 2018 IEEE*, 2018, pp. 1-5.
[6] L. Belkhir and A. Elmeligi, "Assessing ICT global emissions footprint: Trends to 2040 & recommendations," *Journal of Cleaner Production,* vol. 177, pp. 448-463, 2018.
[7] A. Z. Aktas, "Could energy hamper future developments in information and communication technologies (ICT) and knowledge engineering?," *Renewable and Sustainable Energy Reviews,* 2017.
[8] P. Lähdekorpi, M. Hronec, P. Jolma, and J. Moilanen, "Energy efficiency of 5G mobile networks with base station sleep modes," in *Standards for Communications and Networking (CSCN), 2017 IEEE Conference on*, 2017, pp. 163-168.
[9] X. Ge, J. Yang, H. Gharavi, and Y. Sun, "Energy efficiency challenges of 5G small cell networks," *IEEE Communications Magazine,* vol. 55, pp. 184-191, 2017.
[10] R. Bassoli, M. Di Renzo, and F. Granelli, "Analytical energy-efficient planning of 5G cloud radio access network," in *Communications (ICC), 2017 IEEE International Conference on*, 2017, pp. 1-4.
[11] K. Zhang, Y. Mao, S. Leng, Q. Zhao, L. Li, X. Peng, *et al.*, "Energy-efficient offloading for mobile edge computing in 5G heterogeneous networks," *IEEE access,* vol. 4, pp. 5896-5907, 2016.
[12] J. G. Andrews, S. Buzzi, W. Choi, S. V. Hanly, A. Lozano, A. C. K. Soong, *et al.*, "What Will 5G Be?," *IEEE Journal on Selected Areas in Communications,* vol. 32, pp. 1065-1082, 2014.
[13] T. Choi, T. Kim, W. TaverNier, A. Korvala, and J. Pajunpaa, "Agile management of 5G core network based on SDN/NFV technology," in *Information and Communication Technology Convergence (ICTC), 2017 International Conference on*, 2017, pp. 840-844.
[14] H. Hawilo, L. Liao, A. Shami, and V. C. Leung, "NFV/SDN-based vEPC solution in hybrid clouds," in *Communications Conference (MENACOMM), IEEE Middle East and North Africa*, 2018, pp. 1-6.
[15] S. H. Won, M. Mueck, V. Frascolla, J. Kim, G. Destino, A. Pärssinen, *et al.*, "Development of 5G CHAMPION testbeds for 5G services at the 2018 Winter Olympic Games," in *Signal Processing Advances in Wireless Communications (SPAWC), 2017 IEEE 18th International Workshop on*, 2017, pp. 1-5.
[16] V. Q. Rodriguez and F. Guillemin, "Cloud-RAN modeling based on parallel processing," *IEEE Journal on Selected Areas in Communications,* vol. 36, pp. 457-468, 2018.
[17] G. C. Valastro, D. Panno, and S. Riolo, "A SDN/NFV based C-RAN architecture for 5G Mobile Networks," in *2018 International Conference on Selected Topics in Mobile and Wireless Networking (MoWNeT)*, 2018, pp. 1-8.
[18] A. Tzanakaki, M. Anastasopoulos, I. Berberana, D. Syrivelis, P. Flegkas, T. Korakis, *et al.*, "Wireless-optical network convergence: enabling the 5G architecture to support operational and end-user services," *IEEE Communications Magazine,* vol. 55, pp. 184-192, 2017.
[19] M. Riva, H. Donâncio, F. R. Almeida, G. B. Figueiredo, R. I. Tinini, R. M. Cesar Jr, *et al.*, "An Elastic Optical Network-based Architecture for the 5G Fronthaul," in *Simpósio Brasileiro de Redes de Computadores (SBRC)*, 2018.
[20] J. Luo, Q. Chen, and L. Tang, "Reducing Power Consumption by Joint Sleeping Strategy and Power Control in Delay-Aware C-RAN," *IEEE Access,* 2018.
[21] 3GPP, "Releases". [Online]. Available: http://www.3gpp.org/specifications/67-releases. [Accessed: 3 March 2016]
[22] C. Monfreid, "The LTE Network Architecture-A Comprehensive Tutorial," *Alcatel-Lucent White Paper,* 2012.
[23] Alcatel-Lucent. Interworking LTE EPC with W-CDMA Packet Switched Mobile Cores. *Alcatel-Lucent White Paper.* Available: http://www.alcatel-lucent.com
[24] A. Al-Quzweeni, T. E. H. El-Gorashi, L. Nonde, and J. M. H. Elmirghani, "Energy efficient network function virtualization in 5G networks," presented at the 17th International Conference on Transparent Optical Networks (ICTON), 2015.
[25] A. Al-Quzweeni, A. Lawey, T. El-Gorashi, and J. M. Elmirghani, "A framework for energy efficient NFV in 5G networks," in

[26] M. H. Alsharif and R. Nordin, "Evolution towards fifth generation (5G) wireless networks: Current trends and challenges in the deployment of millimetre wave, massive MIMO, and small cells," in *Telecommunication Systems:Modelling, Analysis, Design and Management*, ed: Springer US, 2016, pp. 1-21.

[27] M. Jaber, M. A. Imran, R. Tafazolli, and A. Tukmanov, "5G Backhaul Challenges and Emerging Research Directions: A Survey," *IEEE Access,* vol. 4, pp. 1743-1766, 2016.

[28] P. Chanclou, A. Pizzinat, F. Le Clech, T. L. Reedeker, Y. Lagadec, F. Saliou*, et al.*, "Optical fiber solution for mobile fronthaul to achieve cloud radio access network," ed, 2013, pp. 1-11.

[29] Z. Tayq, "Fronthaul integration and monitoring in 5G networks," Université de Limoges, 2017.

[30] S. Little, "Is microwave backhaul up to the 4G task?," *IEEE microwave magazine,* vol. 10, 2009.

[31] A. Pizzinat, P. Chanclou, F. Saliou, and T. Diallo, "Things you should know about fronthaul," *Journal of Lightwave Technology,* vol. 33, pp. 1077-1083, 2015.

[32] R. Chundury, "Mobile broadband backhaul: Addressing the challenge," *Planning Backhaul Networks, Ericsson Review,* pp. 4-9, 2008.

[33] Alcatel-Lucent. LTE Mobile Transport Evolution. *Alcatel-Lucent White Paper.* Available: http://www.alcatel-lucent.com

[34] R. Kwan and C. Leung, "A survey of scheduling and interference mitigation in LTE," *Journal of Electrical and Computer Engineering,* vol. 2010, p. 1, 2010.

[35] H. G. Myung, "Technical overview of 3GPP LTE," *Polytechnic University of New York,* 2008.

[36] C. Hoymann, W. Chen, J. Montojo, A. Golitschek, C. Koutsimanis, and X. Shen, "Relaying operation in 3GPP LTE: challenges and solutions," *IEEE Communications Magazine,* vol. 50, 2012.

[37] J. Zyren, "Overview of the 3GPP long term evolution physical layer," *White Paper,* 2007.

[38] Anritsu, "LTE Resource Guide", 2015. [Online]. Available: http://www.cs.columbia.edu/6181/hw/anritsu.pdf. [Accessed: 7 May 2018]

[39] R. F. Chisab and C. Shukla, "Performance Evaluation Of 4G-LTE-SCFDMA Scheme Under SUI And ITU Channel Models," *International Journal of Engineering & Technology IJET-IJENS,* vol. 14, 2014.

[40] M. Rinne and O. Tirkkonen, "LTE, the radio technology path towards 4G," *Computer Communications,* vol. 33, pp. 1894-1906, 2010.

[41] A. de la Oliva, J. A. Hernández, D. Larrabeiti, and A. Azcorra, "An overview of the CPRI specification and its application to C-RAN-based LTE scenarios," *IEEE Communications Magazine,* vol. 54, pp. 152-159, 2016.

[42] CPRI Specification V7. 0, Oct 2015 [Online]. Available: http://www.cpri.info/downloads/CPRI_v_7_0_2015-10-09.pdf

[43] J. P. Castro, *The UMTS network and radio access technology*: John Wiley Sons Limited, 2001.

[44] H. Kaaranen, *UMTS networks: architecture, mobility and services*: John Wiley & Sons, 2005.

[45] T. Werthmann, H. Grob-Lipski, and M. Proebster, "Multiplexing gains achieved in pools of baseband computation units in 4G cellular networks," in *Personal Indoor and Mobile Radio Communications (PIMRC), 2013 IEEE 24th International Symposium on*, 2013, pp. 3328-3333.

[46] A. Q. Lawey, T. E. El-Gorashi, and J. M. Elmirghani, "Distributed Energy Efficient Clouds Over Core Networks," *Journal of Lightwave Technology,* vol. 32, pp. 1261-1281, 2014.

[47] A. Q. Lawey, T. El-Gorashi, and J. M. Elmirghani, "Energy efficient cloud content delivery in core networks," in *IEEE Globecom Workshops (GC Wkshps)*, 2013, pp. 420-426.

[48] X. Dong, T. El-Gorashi, and J. M. H. Elmirghani, "Green IP over WDM networks with data centers," *Journal of Lightwave Technology,* vol. 29, pp. 1861-1880, 2011.

[49] N. I. Osman, T. El-Gorashi, L. Krug, and J. M. H. Elmirghani, "Energy-Efficient Future High-Definition TV," *Journal of Lightwave Technology,* vol. 32, pp. 2364-2381, 2014.

[50] H. M. M. Ali, T. E. El-Gorashi, A. Q. Lawey, and J. M. Elmirghani, "Future energy efficient data centers with disaggregated servers," *Journal of Lightwave Technology,* vol. 35, pp. 5361-5380, 2017.

[51] A. M. Al-Salim, T. El-Gorashi, A. Lawey, and J. Elmirghani, "Energy efficient big data networks: impact of volume and variety," *J. Trans. Netw. Serv. Manag.(TNSM),* 2017.

[52] I. Waßmann, D. Versick, and D. Tavangarian, "Energy consumption estimation of virtual machines," in *Proceedings of the 28th Annual ACM Symposium on Applied Computing*, 2013, pp. 1151-1156.

[53] Cisco, "Cisco ASR 5000 Series Product Overview Release 12.0," 2013.

[54] Alcatel-Lucent, "Alcatel-Lucent 9926 Base Band Unit LR13.1.L," 2013.

[55] Intel, "Export Compliance Metrics for Intel® Microprocessors," 2018.

[56] S. Electric, "FTE7502 EPON Optical Network Unit (10G ONU) datasheet". [Online]. Available: http://www.sumitomoelectric.com/onu-fte7502.html. [Accessed: 21 Feb 2015]

[57] Cisco. Cisco ME 4600 Series Optical Line Terminal Data Sheet [Online]. Available: http://www.cisco.com/c/en/us/products/collateral/switches/me-4600-series-multiservice-optical-access-platform/datasheet-c78-730445.html

[58] Alcatel-Lucent. TRDU2x40-08 LTE 3GPP Band 20 LTE FDD Transmit Receive Duplexer Unit – 800 MHz EDD Datasheet [Online].

[59] L. Nonde, T. E. H. El-Gorashi, and J. M. H. Elmirghani, "Energy Efficient Virtual Network Embedding for Cloud Networks," *Journal of Lightwave Technology,* vol. 33, pp. 1828-1849, 2015.

[60] A. Q. Lawey, T. E. H. El-Gorashi, and J. M. H. Elmirghani, "Renewable energy in distributed energy efficient content delivery clouds," presented at the IEEE International Conference on Communications (ICC), 2015.

[61] X. Dong, "On the Energy Efficiency of Physical Topology Design for IP over WDM Networks," *Journal of Lightwave Technology,* vol. 30, pp. 1931-1942, 2012.

[62] X. Dong, T. El-Gorashi, and J. M. H. Elmirghani, "IP Over WDM Networks Employing Renewable Energy Sources," *Journal of Lightwave Technology,* vol. 29, pp. 3-14, 2011.

[63] GreenTouch, "GreenTouch Final Results from Green Meter Research Study", 2015. [Online]. Available: http://www.greentouch.org/index.php?page=greentouch-green-meter-research-study. [Accessed: 7 Jan 2016]

[64] J. Elmirghani, T. Klein, K. Hinton, L. Nonde, A. Lawey, T. El-Gorashi*, et al.*, "GreenTouch GreenMeter core network energy-efficiency improvement measures and optimization," *Journal of Optical Communications and Networking,* vol. 10, pp. A250-A269, 2018.

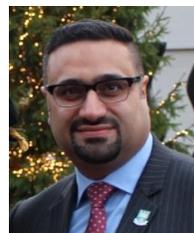

**Ahmed N. Al-Quzweeni** received the B.Sc. and M.Sc. degrees in computer engineering from Al-Nahrain University, Baghdad, Iraq in 2001 and 2004.

He is currently working toward the Ph.D. in Electrical Engineering at the University of Leeds, Leeds, U.K.

From (2009 to 2014) he was assistant lecturer at the department of computer communication in Al-Mansour university college, Baghdad, Iraq.

From (2005 to 2009) he was a mobile core network senior engineer and short message system, intelligent network, PSTN, and billing system team leader at ZTE Corporation for Telecommunication, Iraq branch. His current research interests include energy efficiency in optical and wireless networks, NFV, mobile networks, 5G networks, caching the contents, cloud computing and Internet of Things.

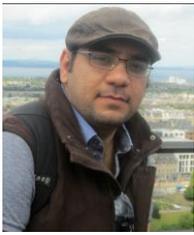

**Ahmed Q. Lawey** received the B.Sc. degree (first-class Honors) in computer engineering from Al-Nahrain University, Iraq, in 2002, the M.Sc. degree (with distinction) in computer engineering from Al-Nahrain University, Iraq, in 2005, and the PhD degree in communication networks from the University of Leeds, UK, in 2015. From 2005 to 2010 he was a core network engineer in ZTE Corporation for Telecommunication, Iraq branch. He is currently a lecturer in communication networks in the School of Electronic and Electrical Engineer, University of Leeds. His current research interests include energy efficiency in optical and wireless networks, big data, cloud computing and Internet of Things.

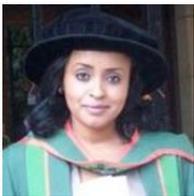

**Taisir E. H. El-Gorashi** received the B.S. degree (first-class Hons.) in electrical and electronic engineering from the University of Khartoum, Khartoum, Sudan, in 2004, the M.Sc. degree (with distinction) in photonic and communication systems from the University of Wales, Swansea, UK, in 2005, and the PhD degree in optical networking from the University of Leeds, Leeds, UK, in 2010. She is currently a Lecturer in optical networks in the School of Electrical and Electronic Engineering, University of Leeds. Previously, she held a Postdoctoral Research post at the University of Leeds (2010–2014), where she focused on the energy efficiency of optical networks investigating the use of renewable energy in core networks, green IP over WDM networks with datacenters, energy efficient physical topology design, energy efficiency of content distribution networks, distributed cloud computing, network virtualization and Big Data. In 2012, she was a BT Research Fellow, where she developed energy efficient hybrid wireless optical broadband access networks and explored the dynamics of TV viewing behavior and program popularity. The energy efficiency techniques developed during her postdoctoral research contributed 3 out of the 8 carefully chosen core network energy efficiency improvement measures recommended by the GreenTouch consortium for every operator network worldwide. Her work led to several invited talks at GreenTouch, Bell Labs, Optical Network Design and Modelling conference, Optical Fiber Communications conference, International Conference on Computer Communications and EU Future Internet Assembly and collaboration with Alcatel Lucent and Huawei.

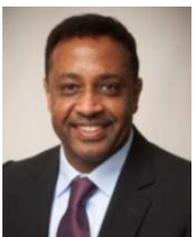

**Jaafar M. H. Elmirghani** (M' 92–SM' 99) is the Director of the Institute of Communication and Power Networks within the School of Electronic and Electrical Engineering, University of Leeds, UK. He joined Leeds in 2007 and prior to that (2000–2007) as chair in optical communications at the University of Wales Swansea he founded, developed and directed the Institute of Advanced Telecommunications and the Technium Digital (TD), a technology incubator/spinoff hub. He has provided outstanding leadership in a number of large research projects at the IAT and TD. He received the BSc in Electrical Engineering, First Class Honours from the University of Khartoum in 1989 and was awarded all 4 prizes in the department for academic distinction. He received the PhD in the synchronization of optical systems and optical receiver design from the University of Huddersfield UK in 1994 and the DSc in Communication Systems and Networks from University of Leeds, UK, in 2014. He has coauthored Photonic switching Technology: Systems and Networks, (Wiley) and has published over 450 papers. He has research interests in optical systems and networks. Prof. Elmirghani is Fellow of the IET, Chartered Engineer, Fellow of the Institute of Physics and Senior Member of IEEE. He was Chairman of IEEE Comsoc Transmission Access and Optical Systems technical committee and was Chairman of IEEE Comsoc Signal Processing and Communications Electronics technical committee, and an editor of IEEE Communications Magazine. He was founding Chair of the Advanced Signal Processing for Communication Symposium which started at IEEE GLOBECOM'99 and has continued since at every ICC and GLOBECOM. Prof. Elmirghani was also founding Chair of the first IEEE ICC/GLOBECOM optical symposium at GLOBECOM'00, the Future Photonic Network Technologies, Architectures and Protocols Symposium. He chaired this Symposium, which continues to date under different names. He was the founding chair of the first Green Track at ICC/GLOBECOM at GLOBECOM 2011, and is Chair of the IEEE Green ICT initiative within the IEEE Technical Activities Board (TAB) Future Directions Committee (FDC), a pan IEEE Societies initiative responsible for Green ICT activities across IEEE, 2012-present. He is and has been on the technical program committee of 34 IEEE ICC/GLOBECOM conferences between 1995 and 2016 including 15 times as Symposium Chair. He has given over 55 invited and keynote talks over the past 8 years. He received the IEEE Communications Society Hal Sobol award, the IEEE Comsoc Chapter Achievement award for excellence in chapter activities (both in international competition in 2005), the University of Wales Swansea Outstanding Research Achievement Award, 2006; and received in international competition: the IEEE Communications Society Signal Processing and Communication Electronics outstanding service award, 2009, a best paper award at IEEE ICC'2013. Related to Green Communications he received (i) the IEEE Comsoc Transmission Access and Optical Systems outstanding Service award 2015 in recognition of "Leadership and Contributions to the Area of Green Communications", (ii) the GreenTouch 1000x award in 2015 for "pioneering research contributions to the field of energy efficiency in telecommunications", (iii) the IET 2016 Premium Award for best paper in IET Optoelectronics and (iv) shared the 2016 Edison Award in the collective disruption category with a team of 6 from GreenTouch for their joint work on the GreenMeter. He is currently an editor of: IET Optoelectronics and Journal of Optical Communications, and was editor of IEEE Communications Surveys and Tutorials and IEEE Journal on Selected Areas in Communications series on Green Communications and Networking. He was Co-Chair of the GreenTouch Wired, Core and Access Networks Working Group, an adviser to the Commonwealth Scholarship Commission, member of the Royal Society International Joint Projects Panel and member of the Engineering and Physical Sciences Research Council (EPSRC) College. He has been awarded in excess of £22 million in grants to date from EPSRC, the EU and industry and has held prestigious fellowships funded by the Royal Society and by BT. He was an IEEE Comsoc Distinguished Lecturer 2013-2016.